\documentclass[onecolumn,prd,nofootinbib,aps,floats,floatfix,amsmath,amssymb,secnumarabic,preprintnumbers,showpacs,notitlepage]{revtex4-1} %
\usepackage[final]{graphicx}
\usepackage{ulem}
\usepackage{amsmath}

\def\beq{\begin{equation}}
\def\eeq{\end{equation}}
\def\bea{\begin{eqnarray}}
\def\eea{\end{eqnarray}}
\def\nn{\nonumber}
\def\roughly#1{\mathrel{\raise.3ex\hbox
{$#1$\kern-.75em\lower1ex\hbox{$\sim$}}}}
\def\lsim{\roughly<}

\def\bd{B^0}

\def\bs{B^0_s}
\def\bsbar{{\bar B}^0_s}
\def\bsmumu{b \to s \mu^+ \mu^-}
\def\bstautau{b \to s \tau^+ \tau^-}

\def\bsll{b \to s \ell^+ \ell^-}
\def\bsnunubar{b \to s \nu {\bar\nu}}
\def\bctaunu{b \to c \tau^- {\bar\nu}}

\def \hc{{\rm h.c.}}

\def \cB{{\cal B}}
\def \cL{{\cal L}}

\def \SM{{\rm SM}}

\def \NP{{\rm NP}}

\def \eff{{\rm eff}}

\def \De{\Delta}
\def \cL{{\cal L}}
\def \oQ{\overline{Q}}
\def \oL{\overline{L}}
\def \si{\sigma}
\def \ga{\gamma}
\def \({\left(}
\def \){\right)}
\def \[{\left[}
\def \]{\right]}
\def \<{\left<}
\def \>{\right>}
\def \l|{\left|}
\def \r|{\right|}

\begin{document}

\title{\boldmath Combined Explanations of the $\bsmumu$ and $\bctaunu$ Anomalies: \\ a General Model Analysis}
\preprint{UdeM-GPP-TH-18-264}
\author{Jacky Kumar}
\email{jkumar@iisermohali.ac.in}
\affiliation{Department of Physics, Indian Institute of Science Education and Research, \\ Mohali, Punjab, 140036 India}
\author{David London}\email{london@lps.umontreal.ca}
\affiliation{Physique des Particules, Universit\'e de Montr\'eal, \\
C.P. 6128, succ. centre-ville, Montr\'eal, QC, Canada H3C 3J7}
\author{Ryoutaro Watanabe}\email{watanabe@lps.umontreal.ca}
\affiliation{Physique des Particules, Universit\'e de Montr\'eal, \\
C.P. 6128, succ. centre-ville, Montr\'eal, QC, Canada H3C 3J7}

\begin{abstract}
There are four models of tree-level new physics (NP) that can
potentially simultaneously explain the $\bsmumu$ and $\bctaunu$
anomalies. They are the $S_3$, $U_3$ and $U_1$ leptoquarks (LQs), and
a triplet of standard-model-like vector bosons ($VB$s). Under the
theoretical assumption that the NP couples predominantly to the third
generation, previous analyses found that, when constraints from other
processes are taken into account, the $S_3$, $U_3$ and $VB$ models
cannot explain the $B$ anomalies, but $U_1$ is viable. In this paper,
we reanalyze these models, but without any assumption about their
couplings. We find that, even in this most general case, $S_3$ and
$U_3$ are excluded. For the $U_1$ model, constraints from the
semileptonic lepton-flavour-violating (LFV) processes $B \to K^{(*)}
\mu^\pm \tau^\mp$, $\tau \to \mu \phi$ and $\Upsilon \to \mu \tau$,
which have been largely ignored previously, are found to be very
important.  Because of the LFV constraints, the pattern of couplings
of the $U_1$ LQ is similar to that obtained with the above theoretical
assumption.  Also, the LFV constraints render unimportant those
constraints obtained using the renormalization group equations. As for
the $VB$ model, it is excluded if the above theoretical assumption is
made due to the additional constraints from $\bs$-$\bsbar$ mixing,
$\tau\to 3\mu$ and $\tau \to \mu \nu {\bar\nu}$. By contrast, we find
a different set of NP couplings that both explains the $\bsmumu$
anomaly and is compatible with all constraints. However, it does not
reproduce the measured values of the $\bctaunu$ anomalies -- it would
be viable only if future measurements find that the central values of
these anomalies are reduced. Even so, this $VB$ model is excluded by
the LHC bounds on high-mass resonant dimuon pairs. This conclusion is
reached without any assumptions about the NP couplings.
\end{abstract}

\maketitle

\section{Introduction}
\label{Sec:Intro}

At the present time, there are a number of measurements of $B$ decays
that are in disagreement with the predictions of the standard model
(SM). These can be separated into two categories:
\begin{enumerate}

\item $\bsmumu$: discrepancies with the SM can be found in several
  observables in $B \to K^* \mu^+\mu^-$ \cite{BK*mumuLHCb1,
    BK*mumuLHCb2, BK*mumuBelle, BK*mumuATLAS, BK*mumuCMS} and $B_s^0
  \to \phi \mu^+ \mu^-$ \cite{BsphimumuLHCb1, BsphimumuLHCb2} decays,
  as well as in the observation of lepton flavour universality (LFU)
  violation in $R_K \equiv \cB(B^+ \to K^+ \mu^+ \mu^-)/{\cal
    B}(B^+ \to K^+ e^+ e^-)$ \cite{RKexpt} and $R_{K^*} \equiv {\cal
    B}(B^0 \to K^{*0} \mu^+ \mu^-)/\cB(B^0 \to K^{*0} e^+ e^-)$
  \cite{RK*expt}.  Following the announcement of the $R_{K^*}$ result,
  several papers performed a combined analysis of the various $\bsll$
  observables \cite{Capdevila:2017bsm, Altmannshofer:2017yso,
    DAmico:2017mtc, Hiller:2017bzc, Geng:2017svp, Ciuchini:2017mik,
    Celis:2017doq, Alok:2017sui}. The general consensus was that the
  discrepancy with the SM is at the level of 4-6$\sigma$ (the range
  reflects the fact that the groups used different ways of treating
  the theoretical hadronic uncertainties). Apart from the size of the
  disagreement, what is particularly intriguing here is that the data
  can all be explained if there is new physics (NP) in $\bsmumu$
  transitions.

\item $\bctaunu$: there are also measurements of LFU violation in
  $R_{D^{(*)}} \equiv \cB(\bar{B} \to D^{(*)} \tau^{-}
  {\bar\nu}_\tau)/\cB(\bar{B} \to D^{(*)} \ell^{-} {\bar\nu}_\ell)$
  ($\ell = e,\mu$) \cite{RD_BaBar, RD_Belle, RD_LHCb,
    Abdesselam:2016xqt} and $R_{J/\psi} \equiv \cB(B_c^+ \to
  J/\psi\tau^+\nu_\tau) / \cB(B_c^+ \to J/\psi\mu^+\nu_\mu)$
  \cite{Aaij:2017tyk}.  Following the measurements of $R_{D^{(*)}}$,
  updated studies of the SM predictions were performed
  \cite{Bernlochner:2017jka, Bigi:2017jbd}. It was found that,
  together, the deviation of the $R_D$ and $R_{D^*}$ measurements from
  the SM predictions is at the 4$\sigma$ level. The discrepancy in
  $R_{J/\psi}$ is 1.7$\sigma$ \cite{Watanabe:2017mip}. These suggest
  the presence of NP in $\bctaunu$ decays.

\end{enumerate}

Much work was done examining NP models that could explain the
$\bsmumu$ or $\bctaunu$ anomalies. One conclusion of these studies was
that the discrepancies can be explained by NP that couples principally
to left-handed (LH) particles, i.e., its interactions are of the form
$(V-A) \times (V-A)$. In Ref.~\cite{RKRD}, it was pointed out that, if
the NP couples to LH particles, one can relate the neutral-current
$\bsmumu$ and charged-current $\bctaunu$ transitions using the SM
$SU(2)_L$ symmetry. That is, it is possible to find a NP model that
can simultaneously explain the $\bsmumu$ and $\bctaunu$ anomalies.

Following this observation, there was a great deal of activity
examining various aspects of simultaneous explanations of both
$B$-decay anomalies \cite{AGC, Isidori, CCO, Bauer:2015knc, U3LQ,
  Barbieri:2015yvd, Boucenna:2016wpr, Das:2016vkr, Feruglio:2016gvd,
  Boucenna:2016qad, Becirevic:2016yqi, Sahoo:2016pet, RKRDmodels,
  Barbieri:2016las, Bordone:2017bld, Chen:2017hir, Megias:2017ove,
  Crivellin:2017zlb, Altmannshofer:2017poe, Alok:2017jaf,
  Feruglio:2017rjo, Matsuzaki:2017bpp, Dorsner:2017ufx, Zurich,
  Choudhury:2017qyt, Assad:2017iib, DiLuzio:2017vat, Calibbi:2017qbu,
  Chauhan:2017uil, Bordone:2017anc, Choudhury:2017ijp, Fuyuto:2017sys,
  Barbieri:2017tuq, DAmbrosio:2017wis, Blanke:2018sro,
  Azatov:2018kzb}. Many of these papers studied specific models. It
was found that, if one insists on LH NP that contributes to both
$\bsmumu$ and $\bctaunu$ at tree level, there are only four types of
NP models. There are three leptoquark (LQ) models: (i) $S_3$,
containing an $SU(2)_L$-triplet scalar LQ, (ii) $U_3$, an
$SU(2)_L$-triplet vector LQ, and (iii) $U_1$, an $SU(2)_L$-singlet
vector LQ. And there is the $VB$ model, which contains SM-like LH $W'$
and $Z'$ vector bosons.

In Refs.~\cite{RKRDmodels} and \cite{Zurich}, all four models were
studied, taking into account not only the $\bsmumu$ and $\bctaunu$
data, but also constraints from other processes to which the
particular NP contributes. Of the two anomalies, the NP effect in
$\bctaunu$ is larger (in absolute size, not relative to the SM),
simply because the process is tree level in the SM. Of the four
particles involved in this transition, three of them belong to the
third generation, with the fourth in the second generation. It is then
quite natural to assume that the NP couples predominantly to the
third generation, with the couplings involving the second generation
subdominant.

This is the assumption made in Refs.~\cite{RKRDmodels} and
\cite{Zurich}, though its implementation differs in the two papers. In
Ref.~\cite{RKRDmodels}, it is assumed that the NP couples only to the
third generation in the weak basis. The couplings to the second
generation are induced when one transforms to the mass basis. Since
the mixing angles involved in this transformation are small, the
couplings in the mass basis obey a hierarchy $|c_{22}| < |c_{23}|,
|c_{32}| < |c_{33}|$, where the indices indicate the generations.  In
Ref.~\cite{Zurich}, an $U(2)_q \times U(2)_\ell$ flavour symmetry is
imposed, so that the NP couples only to the third generation (in the
mass basis). The couplings to the second generation are generated by
symmetry-breaking terms due to spurions.  Here too, the couplings obey
the above hierarchy.

We note in passing that the assumption of NP coupling only to the
third generation in the weak basis was quite popular. It was applied
in a number of papers, on a variety of subjects -- model-independent
analyses, specific models, and UV completions of the $VB$ and $U_1$
models.

In both analyses the $S_3$, $U_3$ and $VB$ models were ruled out; only
the $U_1$ model was a viable candidate for explaining all the
$B$-decay anomalies.  But this raises the question: to what extent do
these conclusions depend on the assumption regarding the NP couplings?
While the idea of NP coupling principally to the third generation is
attractive theoretically, it is not the only possibility. If one
relaxes this assumption, so that the couplings involving the second
generation are no longer subdominant, could we find $S_3$, $U_3$ or
$VB$ models that can account for the $\bsmumu$ and $\bctaunu$ data?
How does the $U_1$ model change in this case?

This is the issue we address in this paper. We focus separately on the
LQ and $VB$ models. In both cases, we work solely in the mass basis.
For simplicity, we assume that the NP couplings involving the first
generation leptons and down-type quarks are negligible. (This allows
us to focus on the second and third generations, which participate in
$\bsmumu$ and $\bctaunu$.) Our idea is simply to establish what sizes
of NP couplings are required by the data.

We show that the $S_3$ and $U_3$ LQ models cannot explain the
$B$-decay anomalies, even if only constraints from the anomalies and
$B \to K^{(*)} \nu {\bar\nu}$ are taken into account. On the other
hand, the $U_1$ model is a viable explanation. If only these
constraints are imposed, the couplings can take a great many values.
However, when one includes the constraints from semileptonic processes
that exhibit lepton flavour violation (LFV), namely $B \to K^{(*)}
\mu^\pm \tau^\mp$, $\tau \to \mu \phi$ and $\Upsilon \to \mu\tau$, one
finds that the region of allowed couplings is greatly reduced. It is
similar (though somewhat larger) to that found when the NP couples
predominantly to the third generation. In other words, the data
actually point in this direction; no theoretical assumptions are
necessary.

When one evolves the full Lagrangian from the NP scale down to low
enrgies using the one-loop renormalization group equations (RGEs), one
generates new contributions to a variety of operators. It has been
argued \cite{Feruglio:2016gvd,Feruglio:2017rjo} that the additional
constraints due to these new effects lead to an important reduction in
the allowed space of couplings. In this paper, we point out that these
RGE constraints are not rigorous. More importantly, we show that, if
the absolute value of all couplings is taken to be $\le 1$, so that
they remain perturbative, the LFV constraints are much more stringent
than the RGE constraints.

In the case of the $VB$ model, the result is different. In this model,
there are also tree-level contributions to $\bs$-$\bsbar$ mixing,
$\tau\to 3\mu$, $\tau \to \mu \nu {\bar\nu}$ and $D^0$-${\bar D}^0$
mixing, and these lead to additional severe constraints on the
couplings. In particular, the $Z' \mu^\pm \tau^\mp$ coupling must be
very small. But if the NP couples principally to the third generation,
this coupling is always rather sizeable, so that this $VB$ model is
ruled out.

On the other hand, in this more general case, we find a set of
couplings that both explains the $\bsmumu$ anomaly and is compatible
with all constraints. However, it does not reproduce the measured
values of the $\bctaunu$ anomalies. There is an enhancement of
$R_{D^{(*)}}$, but it is smaller than what is observed.  If future
measurements of $R_{D^{(*)}}$ confirm the present measurements, then
the $VB$ modell will be ruled out. Still, if it is found that the
central values of $R_{D^{(*)}}$ are reduced, the $VB$ model could be
an explanation of both anomalies.  For this reason, as far as the
anomalies are concerned, we refer to the model as semi-viable.

Unfortunately, with this set of couplings, the predicted rate for the
production of high-mass resonant dimuon pairs at the LHC is larger
than the limits placed by ATLAS and CMS. We note that this constraint
can be evaded by adding additional, invisible decays of the $Z'$.  If
this possibility is not realized, we find that, in the end, the $VB$
model is excluded. However, we stress that this is not the result of
any assumption about the NP couplings. Rather, it is found simply by
taking into account all the flavour constraints and the bound from the
LHC dimuon search.

We begin in Sec.~\ref{Sec:Obs} with a summary of the observables
necessary for this study. In Sec.~\ref{Sec:LQmodels}, we examine the
leptoquark models. We show that the $S_3$ and $U_3$ models are ruled
out, determine the pattern of couplings necessary for the $U_1$ model
to explain the $B$ anomalies, and tabulate the predictions of this
model for other processes. A similar study of the $VB$ model is
carried out in Sec.~\ref{Sec:VBmodel}. We show that the model is
excluded if the $Z' \mu^\pm \tau^\mp$ coupling is sizeable. We also
demonstrate that, if this coupling is very small, the model is
semi-viable but also leads to a disagreement with the LHC bounds on
the production of high-mass resonant dimuon pairs. We conclude in
Sec.~\ref{Sec:Conclusions}.

\section{Observables}
\label{Sec:Obs}

The $B$ anomalies involve the decays $\bsmumu$ and $\bctaunu$, both
semileptonic processes with two quarks and two leptons ($2q2\ell$).
There are two $2q2\ell$ operators that are invariant under the full
$SU(3)_C \times SU(2)_L \times U(1)_Y$ gauge group. In the mass basis,
they are given by\footnote{These operators are also used in the SM
  Effective Field Theory, see, for example,
  Refs.~\cite{Aebischer:2015fzz, Dedes:2017zog}.}
\beq
\mathcal L_{\rm NP} =
\frac{G_1^{ijkl}}{\Lambda_{\rm NP}^2} ({\bar Q}_{iL} \gamma_\mu Q_{jL}) ({\bar L}_{kL} \gamma^\mu L_{lL}) 
+
\frac{G_3^{ijkl}}{\Lambda_{\rm NP}^2} ({\bar Q}_{iL} \gamma_\mu \sigma^I Q_{jL}) ({\bar L}_{kL} \gamma^\mu \sigma^I L_{lL}) ~, 
\label{2q2lops}
\eeq
where $\sigma^I$ ($I=1,2,3$) are the Pauli matrices, and $Q_L$ and
$L_L$ are left-handed quark and lepton doublets, defined as
\beq
\label{NPcouplings}
 Q_L = 
 \begin{pmatrix} V^\dag\, u_L \\ d_L \end{pmatrix}  ~~,~~~~ 
 L_L = 
 \begin{pmatrix} \nu_L \\ \ell_L \end{pmatrix}  ~. 
\eeq
Here $V$ denotes the Cabibbo-Kobayashi-Maskawa (CKM) matrix. NP models
that simultaneously explain the two $B$ anomalies are distinguished by
their $G_1$ and $G_3$ factors.

NP models that can explain the $\bsmumu$ and $\bctaunu$ anomalies must
contribute to these decays. From the above, we see that they can
potentially contribute to other $2q2\ell$ processes. A complete
analysis of any possible NP model must therefore consider constraints
from all $2q2\ell$ observables.

These observables can be separated into neutral-current (NC) and
charged-current (CC) processes. The NC observables can themselves be
separated into four types: lepton-flavour-conserving (LFC) branching
ratios (BRs), lepton-flavour-universality-violating (LFUV) ratios of
BRs, lepton-flavour-violating (LFV) decays, and invisible decays. The
full list of these obserables that have been measured is
\cite{Aloni:2017eny}
\bea
{\hbox{LFC BRs}} &:& \Upsilon(nS) \to \ell^+ \ell^-; J/\psi
\to \mu^+ \mu^-; \phi \to \mu^+ \mu^-; \bs \to \mu^+ \mu^-;
\bs \to \phi\mu^+ \mu^-; B \to K^{(*)} \mu^+ \mu^- ~, \nn\\
{\hbox{LFUV ratios}} &:& R_{\Upsilon(nS)}^{\ell/\ell'}; R_{J/\psi}^{\mu/e}; R_{\phi}^{\mu/e}; R_{B \to
  K^{(*)}}^{e/\mu} ~, \nn\\
{\hbox{LFV decays}} &:& \Upsilon(nS) \to \mu^\pm \tau^\mp; J/\psi \to
\mu^\pm \tau^\mp; \tau \to \mu \phi; B \to K^{(*)} \mu^\pm \tau^\mp ~, \nn\\
{\hbox{Invisible}} &:& \Upsilon(nS) \to \nu {\bar\nu}; J/\psi \to \nu
{\bar\nu}; \phi \to \nu {\bar\nu}; \bs \to \phi \nu {\bar\nu}; B \to
K^{(*)} \nu {\bar\nu} ~.
\eea
In the LFC BRs, $\ell = \tau, \mu$, while in the LFUV ratios,
$\ell/\ell' = \tau/\mu, \tau/e, \mu/e$. The CC observables come in two
types: LFC BRs and LFUV ratios. These are
\bea
{\hbox{LFC BRs}} &:& B_c^+ \to J/\psi \ell^+ \nu_\ell; {\bar
  B} \to D^{(*)} \ell^- {\bar\nu}_\ell; D_s^+ \to \ell^+ \nu_\ell; \nn\\
&& \hskip1truecm D^+ \to {\bar K}^0 \mu^+ \nu_\mu, D^0 \to K^{(*)-}\mu^+ \nu_\mu ~, \nn\\
{\hbox{LFUV ratios}} &:& R_{J/\psi}^{\tau/\mu}; R_{D^{(*)}}^{\tau/\ell}; R_{D^{(*)}}^{\mu/e};
R_{D_s}^{\tau/\mu}; R_{D^+ \to {\bar K}^0}^{\mu/e}, R_{D^0 \to {\bar
    K}^{(*)+}}^{\mu/e} ~.
\eea
In the LFC BRs, $\ell = \tau, \mu$. In the above,
$R_{\Upsilon(nS)}^{\tau/\mu} \equiv {\cal B}(\Upsilon(nS) \to \tau^+
\tau^-)/\cB(\Upsilon(nS) \to \mu^+ \mu^-)$. The other LFUV ratios are
defined similarly. There are additional $2q2\ell$ observables, such as
$\cB(B \to K^* \tau^+ \tau^-)$, LFUV in $B^- \to \ell^- \nu_\ell$,
etc., that have not yet been measured, but are likely to be in the
near future. These will be included in our discussion of predictions
(Sec.~\ref{Sec:LQPredictions}).

Ideally, analyses of NP models would include constraints from all of
these observables.  However, most analyses focus only on a subset of
these observables, which we call the ``minimal constraints.'' These
include observables that involve the decays $\bsmumu$ ($B \to K^{(*)}
\mu^+\mu^-$, $\bs \to \phi \mu^+ \mu^-$, $\bs \to \mu^+ \mu^-$,
$R_{K^{(*)}}$), $\bctaunu$ ($R_{D^{(*)}}$, $R_{J/\psi}$) and
$\bsnunubar$ ($B \to K^{(*)} \nu {\bar\nu}$, $\bs \to \phi \nu
   {\bar\nu}$). The effective Hamiltonians for these processes are
\bea
H_{\rm eff}(b \to s \mu^+ \mu^-) & = & - {\alpha G_F \over \sqrt 2 \pi} V_{tb} V_{ts}^*\,
 \left[ C_9^{\mu\mu}\, \left(\bar s_L \gamma^\mu b_L \right) \left( \bar\mu \gamma_\mu \mu \right) 
+~C_{10}^{\mu\mu}\, \left(\bar s_L \gamma^\mu b_L \right) \left( \bar\mu \gamma_\mu \gamma^5 \mu \right) \right] ~, \nn\\
H_{\rm eff}(b \to c \ell_i\bar\nu_j) & = & {4 G_F \over \sqrt 2} V_{cb} C_V^{ij}
\left( \bar c_L \gamma^\mu b_L \right) \left(\bar\ell_{iL} \gamma_\mu \nu_{jL} \right) ~, \nn\\
H_{\rm eff}(b \to s \nu_i\bar\nu_j) & = & - {\alpha G_F \over \sqrt 2 \pi} V_{tb} V_{ts}^*\,
C_L^{ij}\, \left(\bar s_L \gamma^\mu b_L \right) \left( \bar\nu_i \gamma_\mu (1-\gamma^5) \nu_j \right) ~,
\eea
where the Wilson coefficients include both the SM and NP
contributions: $C_X = C_X ({\rm SM}) + C_X ({\rm NP})$. These NP
contributions are given by
\bea
\label{NPWCs}
C_9^{\mu\mu}({\rm NP}) = -C_{10}^{\mu\mu}({\rm NP}) &=& 
\frac{\pi}{\sqrt{2} \alpha G_F V_{tb}V_{ts}^*} \,
\frac{(G_1 + G_3)^{b s \mu \mu}}{M^2_{\rm NP}} ~, \nn\\
C_V^{ij}({\rm NP}) &=& - \frac{1}{2 \sqrt{2} G_F V_{cb}} \, \frac{2 (V G_3)^{b c i j}}{M^2_{\rm NP}} ~, \nn\\
C_L^{ij}({\rm NP}) &=& \frac{\pi}{\sqrt{2} \alpha G_F V_{tb}V_{ts}^*} \, \frac{(G_1 - G_3)^{b s i j}}{M^2_{\rm NP}} ~.
\eea

Consider now the other observables. For all $2q2\ell$ processes, the
NP contributes at tree level. This contribution can be significant if
the SM contribution to the process is suppressed. This is the case for
$\bsmumu$ (loop level in the SM) and $\bctaunu$ (the SM amplitude
involves the CKM matrix element $V_{cb} \simeq 0.04$). However, if the
SM contribution is unsuppressed, then it dominates the NP
contribution.  This occurs in all NC observables in which there is
neither quark nor lepton flavour violation, namely the decays of
$\Upsilon(nS)$, $J/\psi$ and $\phi$ to $l^+ l^-$ or $\nu {\bar\nu}$.
It also applies to CC observables governed by the transition $c \to s
l \nu$ ($D_s^+ \to l^+ \nu_l$, $D^+ \to {\bar K}^0 l^+ \nu_l$, $D^0
\to K^{(*)-} l^+ \nu_l$), for which $V_{cs} \simeq 1$. For all of
these observables, their constraints on the LQ couplings are extremely
weak and need not be taken into account.

This leaves only the four LFV observables that can put important
constraints on the NP models:
\begin{itemize}

\item $B \to K^{(*)} \mu^\pm \tau^\mp$: for the final state $\mu^-
  \tau^+$ we have
\beq
C_9^{b s \mu \tau}({\rm NP}) = -C_{10}^{b s \mu \tau}({\rm NP}) = 
-\frac{\pi}{\sqrt{2} \alpha G_F V_{tb}V_{ts}^*} \,
\frac{(G_1 + G_3)^{b s \mu \tau}}{M^2_{\rm NP}} ~. \nn
\eeq
For the final state $\tau^- \mu^+$ the NP Wilson coefficient $C_9^{b s
  \tau \mu}({\rm NP}) = -C_{10}^{b s \tau \mu}({\rm NP})$ is found by
replacing $b s \mu \tau \to b s \tau \mu$. The branching
ratios for $B \to K^{(*)} \mu^- \tau^+$ are given in Ref.~\cite{CCO}
and are repeated below:
\bea
 B_{\mu^-\tau^+}^{B\to K} 
 &=& 
 \left( (9.6 \pm 1.0) |C_{9}^{bs\mu\tau}({\rm NP})|^2 + (10.0 \pm 1.3) |C_{10}^{bs\mu\tau}({\rm NP})|^2 \right) \times 10^{-9} ~, \nn\\
 B_{\mu^-\tau^+}^{B\to K^*} 
 &=&  
 \left( (19.4 \pm 2.9) |C_{9}^{bs\mu\tau}({\rm NP})|^2 + (18.1 \pm 2.6) |C_{10}^{bs\mu\tau}({\rm NP})|^2 \right) \times 10^{-9} ~.
\eea
The branching ratios for $B \to K^{(*)} \tau^- \mu^+$ are given by
replacing $bs\mu\tau$ with $bs\tau\mu$.

\item $\tau \to \mu \phi$:
\beq
C_9^{s s \tau \mu}({\rm NP}) = -C_{10}^{s s \mu \tau}({\rm NP}) = 
-\frac{\pi}{\sqrt{2} \alpha G_F V_{tb}V_{ts}^*} \,
\frac{(G_1 + G_3)^{s s \mu \tau}}{M^2_{\rm NP}} ~. \nn
\eeq
The branching ratio is
\beq
 B_{\tau^+\mu^+}^\phi 
 = {f_{\phi}^2 m_{\tau}^3 \over 128\pi\,\Gamma_\tau} (1 - r_\tau^{-1})^2 (1 + 2r_\tau^{-1}) 
\Big[ |(G_1 + G_3)^{s s \mu \tau}|^2 + |(G_1 + G_3)^{s s \tau \mu}|^2 \Big] ~,
\eeq
where $r_\tau \equiv m_\tau^2/m_\phi^2$ and $f_{\phi} = (238 \pm
3)\,{\rm MeV}$ \cite{Chakraborty:2017hry}.

\item $\Upsilon(nS) \to \mu^\pm \tau^\mp$: the branching ratio is
\beq
 B_{\tau\mu}^{\Upsilon (nS)}
 = {f_{\Upsilon(nS)}^2 m_{\Upsilon(nS)}^3 \over 48\pi\, \Gamma_{\Upsilon (nS)} \, M^4_{\rm NP}} (2 + r'_\tau) (1-r'_\tau)^2 
 \Big[ |(G_1 + G_3)^{b b \mu \tau}|^2 + |(G_1 + G_3)^{b b \tau \mu}|^2 \Big]  ~, 
 \label{eq:UpsilonLFV}
\eeq
where $r'_\tau \equiv m_\tau^2/m_{\Upsilon (nS)}^2$, $f_{\Upsilon(1S)}
= (700 \pm 16)\,{\rm MeV}$, $ f_{\Upsilon(2S)} = (496 \pm 21)\,{\rm
  MeV}$, and $f_{\Upsilon(3S)} = (430 \pm 21)\,{\rm
  MeV}$~\cite{RKRDmodels}.

\item $J/\psi \to \mu^\pm \tau^\mp$: the branching ratio is obtained
  from Eq.~\eqref{eq:UpsilonLFV} by replacing $\Upsilon \to J/\psi$
  and $(G_1 + G_3)^{bb\ell\ell'} \to [V(G_1 -
    G_3)V^\dagger]^{cc\ell\ell'}$, with $f_{J/\psi} = (401 \pm
  46)\,{\rm MeV}$ \cite{Becirevic:2013bsa}.

\end{itemize}

Above, we identified the $2q2\ell$ observables that can significantly
constrain the NP models. We list these observables, along with their
present measured values or constraints, in Table \ref{tab:obs_meas}.

Some comments concerning the entries in the Table may be useful: 
\begin{itemize}

\item A fit to all $\bsmumu$ data ($B \to K^{(*)} \mu^+\mu^-$, $\bs
  \to \phi \mu^+ \mu^-$, $\bs \to \mu^+ \mu^-$, $R_{K^{(*)}}$) was
  done in Ref.~\cite{Alok:2017sui}, leading to the constraint on
  $C_9^{\mu\mu}({\rm LQ}) = -C_{10}^{\mu\mu}({\rm LQ})$ given in the
  Table. 

\item Similarly, the analysis of $B \to K^{(*)} \nu {\bar\nu}$ decays
  done in Ref.~\cite{Alok:2017jgr} leads to the constraint on
  $C_L^{ij}({\rm LQ})$ given in the Table. There is also an upper
  limit on $\cB(\bs \to \phi \nu {\bar\nu})$, but it is much
  weaker than that of $\cB(B \to K^{(*)} \nu {\bar\nu})$.

\item The results of the measurements of LFV processes are usually
  given in terms of 90\% C.L. upper limits (ULs) on the branching
  ratios.  For certain measurements ($B^+ \to K^+ \tau^- \mu^+$, $B^+
  \to K^+ \tau^+ \mu^-$, $\Upsilon(2S) \to \mu^\pm \tau^\mp$) the
  actual central values and errors are given, in addition to the UL.
  These are extremely useful, as they can be included in a fit. For
  other measurements ($\tau \to \mu \phi$, $J/\psi \to \mu^\pm
  \tau^\mp$), only the UL is given. In order to include these
  measurements in a fit, we convert the ULs to a branching ratio of $0
  \pm {\rm UL}/1.5$.

\item Other analyses combine $\cB(B \to K \tau^- \mu^+)$ and
  $\cB(B \to K \tau^+ \mu^-)$. However, in the case of LQ models,
  this is not correct, as the two decays involve different couplings.

\item As we describe later, in this paper we assume that the NP does
  not couple significantly to the first-generation down-type quarks.
  However, it does couple to first-generation up-type quarks via the
  CKM matrix [Eq.~(\ref{NPcouplings})]. As a result, there is an
  additional LFV process to which the NP contributes at tree level:
  $\tau \to \mu \rho^0$. Experimentally, it is found that ${\cal
    B}(\tau \to \mu \rho^0) < 1.2 \times 10^{-8}$ (90\% C.L.)
  \cite{Miyazaki:2011xe}, which is stronger than the other upper
  limits in the Table.  This said, it can be shown that the NP
  contribution to $\tau \to \mu \rho^0$ is $|V_{us}|^2 \simeq 0.05$
  times that to $\tau \to \mu \phi$. As a result, the constraint from
  $\tau \to \mu \rho^0$ is much weaker than that from $\tau \to \mu
  \phi$, and for this reason this LFV process is not included in the
  Table.

\end{itemize}

\begin{table*}[t] 
\begin{center}
\begin{tabular}{|c|c|}
 \hline\hline 
Observable & Measurement or Constraint \\
\hline
minimal & \\
\hline
$\bsmumu$ (all) & $C_9^{\mu\mu}({\rm LQ}) = -C_{10}^{\mu\mu}({\rm LQ}) = -0.68 \pm 0.12$ \cite{Alok:2017sui} \\
$R_{D^*}^{\tau/\ell}/(R_{D^*}^{\tau/\ell})_\SM$ & $1.18 \pm 0.06$ \cite{RD_BaBar, RD_Belle, RD_LHCb,Abdesselam:2016xqt} \\
$R_{D}^{\tau/\ell}/(R_{D}^{\tau/\ell})_\SM$ & $1.36 \pm 0.15$ \cite{RD_BaBar, RD_Belle, RD_LHCb,Abdesselam:2016xqt} \\
$R_{D^*}^{e/\mu}/(R_{D^*}^{e/\mu})_\SM$ & $1.04 \pm 0.05$ \cite{Abdesselam:2017kjf} \\
$R_{J/\psi}^{\tau/\mu}/(R_{J/\psi}^{\tau/\mu})_\SM$ & $2.51 \pm 0.97$ \cite{Aaij:2017tyk} \\
$\cB(B \to K^{(*)} \nu {\bar\nu})/\cB(B \to K^{(*)} \nu {\bar\nu})_\SM$ & 
        $-13 \sum_{i=1}^3 {\rm Re}[C_L^{ii}({\rm LQ})] + \sum_{i,j=1}^3 |C_L^{ij}({\rm LQ})|^2 \le 248$ \cite{Alok:2017jgr} \\
\hline
LFV & \\
\hline
$\cB(B^+ \to K^+ \tau^- \mu^+)$ & $(0.8 \pm 1.7) \times 10^{-5}$ ~;~~ $< 4.5 \times 10^{-5}$ (90\% C.L.)  \cite{Lees:2012zz} \\
$\cB(B^+ \to K^+ \tau^+ \mu^-)$ & $(-0.4 \pm 1.2) \times 10^{-5}$ ~;~~ $< 2.8 \times 10^{-5}$ (90\% C.L.) \cite{Lees:2012zz} \\
$\cB(\Upsilon(2S) \to \mu^\pm \tau^\mp)$ & $(0.2 \pm 1.5 \pm 1.3) \times 10^{-6}$ ~;~~ $< 3.3 \times 10^{-6}$ (90\% C.L.) \cite{Lees:2010jk} \\
$\cB(\tau \to \mu \phi)$ & $< 8.4 \times 10^{-8}$ (90\% C.L.) \cite{Miyazaki:2011xe} \\
$\cB(J/\psi \to \mu^\pm \tau^\mp)$ & $< 2.0 \times 10^{-6}$ (90\% C.L.) \cite{Ablikim:2004nn} \\
 \hline\hline 
\end{tabular}
\end{center}
\caption{Measured values or constraints of the $2q2\ell$ observables
  that can significantly constrain the NP models.}
\label{tab:obs_meas}
\end{table*}

In LQ models, the only NP contributions are to the $2q2\ell$
observables described above. On the other hand, in the $VB$ model,
there are also tree-level contributions to four-quark and four-lepton
observables. The five additional observables that yield important
constraints on the $VB$ model are $\bs$-$\bsbar$ mixing, neutrino
trident production, $\tau \to 3\mu$, $\tau \to \mu \nu {\bar\nu}$ and
$D^0$-${\bar D}^0$ mixing.  These will be discussed in more detail in
Sec.~\ref{Addobs}.

\section{Leptoquark models}
\label{Sec:LQmodels}

There are three types of leptoquarks that contribute to both $\bsmumu$
and $\bctaunu$. They are (i) an $SU(2)_L$-triplet scalar LQ ($S_3$)
[$({\bf 3},{\bf 3},-2/3)$], (ii) an $SU(2)_L$-triplet vector LQ
($U_3$) [$({\bf 3},{\bf 3},4/3)$], and (iii) an $SU(2)_L$-singlet
vector LQ ($U_1$) [$({\bf 3},{\bf 1},4/3)$]. In the mass basis,
their interaction Lagrangians are given by \cite{RDLQs}
\bea
\De\cL_{S_3} &=& h^{S_3}_{ij} \(\oQ_{iL}\si^I i\sigma^2 L^c_{j L}\)S^{I}_{3} + \hc, \nn\\
\De\cL_{U_3} &=& h^{U_3}_{ij} \(\oQ_{i L}~\ga^\mu~\si^I L_{j L}\)U^{I}_{3\mu} ~+~ \hc, \nn\\
\De\cL_{U_1} &=& h^{U_1}_{ij} \(\oQ_{i L}~\ga^\mu~L_{j L}\) U_{1\mu} + \hc
\eea
Note that the $S_3$ coupling violates fermion number, while those of
$U_3$ and $U_1$ are fermion-number conserving.  When the heavy LQ is
integrated out, we obtain the following effective Lagrangians:
\bea
\cL^\eff_{S_3} &=& \frac{h_{ik} h_{jl}^*}{4 M^2_{\rm LQ}}\[3\(\oQ_{iL}
\ga^\mu Q_{jL}\)\(\oL_{kL}\ga_\mu L_{lL}\) + \(\oQ_{iL}\ga^\mu\si^I Q_{jL}\)\(\oL_{kL}\ga_\mu\si
^I L_{lL}\)\] ~, \nn\\
\cL^\eff_{U_3} &=& -~\frac{h_{il} h_{jk}^*}{2 M^2_{\rm LQ}}\[3\(\oQ_{iL}\ga^
\mu Q_{jL}\)\(\oL_{kL}\ga^\mu L_{lL}\) - \(\oQ_{iL}\ga^\mu\si^I
Q_{jL}\)\(\oL_{kL}\ga_\mu\si^I L_{lL}\)\] ~, \nn\\
\cL^\eff_{U_1} &=& -~\frac{h_{il} h_{jk}^*}{2 M^2_{\rm LQ}}\[\(\oQ_{iL}\ga^\mu
Q_{jL}\)\(\oL_{kL}\ga^\mu L_{lL}\) + \(\oQ_{iL}\ga^\mu\si^I Q_{
jL}\)\(\oL_{kL}\ga_\mu\si^I L_{lL}\)\] ~.
\eea
Comparing to Eq.~(\ref{2q2lops}), we see that $G_1^{ijkl}$ is replaced
by a constant $g_1$ times the product of two LQ couplings $h~h^*$, and
similarly for $G_3^{ijkl}$. Note that, for the $S_3$ model, the quarks
are coupled to the opposite leptons than in the $U_3$ and $U_1$
models. This is due to the fact that the couplings violate ($S_3$) or
conserve ($U_3$ and $U_1$) fermion number, and is relevant only for
lepton-flavour-violating processes. In the above, we have suppressed
the LQ model labels on the couplings.  The models are distinguished by
their relative weighting of the two operators, $g_1$ and $g_3$. These
are
\bea
\label{g1g2}
S_3 &:& g_1 = 3 g_3 = \frac34 ~, \nn\\
U_3 &:& g_1 = -3 g_3 = -\frac32 ~, \nn\\
U_1 &:& g_1 = g_3 = -\frac12 ~.
\label{LQgs}
\eea

In this paper, we take the couplings to be real. In addition, since
the $\bsmumu$ and $\bctaunu$ anomalies involve only the second and
third generations, for simplicity we assume that the LQ couplings to
the first-generation leptons and down-type quarks are negligible in
the mass basis\footnote{For discussions of processes that are affected
  if there are also nonzero first-generation couplings, see
  Refs.~\cite{Crivellin:2017dsk, Bobeth:2017ecx}, for example.}.
(Even so, they couple to first-generation up-type quarks via the CKM
matrix, see Eq.~(\ref{NPcouplings}).)

In the following subsections, we confront the three LQ models with the
data. For each of the models, we aim to answer two questions. Can the
model explain the $B$-decay anomalies? If so, taking into account all
constraints from $2q2\ell$ observables, what ranges of couplings are
allowed?

\subsection{$S_3$ and $U_3$ LQs}
\label{S3U3LQs}

For both the $S_3$ and $U_3$ LQ models, we perform a fit to the data
using only the 6 minimal constraints of Table \ref{tab:obs_meas} and
setting $M_{\rm LQ} = 1$ TeV. The theoretical parameters are the 4
couplings $h_{22}$, $h_{23}$, $h_{32}$ and $h_{33}$, so that the
number of degrees of freedom (d.o.f.) is 2.

In the SM, $\chi^2_{\SM} = 52$. We find $\chi^2_{min,\SM + S_3} = 15$,
so the addition of the $S_3$ LQ does indeed improve things. On the
other hand, the $\chi^2_{min}/{\rm d.o.f.} = 7.5$. An acceptable fit
has $\chi^2_{min}/{\rm d.o.f.} \simeq 1$, so that, even with the
addition of the NP, the fit is still very poor. Thus, the $S_3$ LQ
model cannot explain the $B$-decay anomalies. (In
Ref.~\cite{Dorsner:2017ufx}, the $S_3$ LQ was allowed to couple to
both the second and third generations, and the same result was found.)

The analysis of the $U_3$ LQ model is similar. The fit to the 6
minimal constraints yields $\chi^2_{min,\SM + U_3} = 20$, or
$\chi^2_{min}/{\rm d.o.f.} = 10$. Here too the fit is very poor: the
$B$-decay anomalies cannot be explained in the $U_3$ LQ model either.

For both LQs we can understand why this is so. The constraint from the
$\bsmumu$ data implies $(g_1 + g_3) \, h_{32} h_{22} = 0.0011 \pm
0.0002$ for $M_{\rm LQ} = 1$ TeV, while that from $R_{D^{(*)}}$ leads
to $2 g_3 \, h_{33} h_{23} = -0.14 \pm 0.04$. There are several NP
contributions to $B \to K^{(*)} \nu {\bar\nu}$, leading to different
flavours of the final-state neutrinos. However, the most important
ones are those that lead to processes that also appear in the SM. The
reason is that, due to SM-NP interference, there are linear NP terms
in the matrix element. There are two possibilities for the neutrinos:
$\nu_\mu {\bar\nu}_\mu$ and $\nu_\tau {\bar\nu}_\tau$, whose NP
contributions involve $h_{32} h_{22}$ and $h_{33} h_{23}$,
respectively. However, from the above constraints we have $|h_{32}
h_{22}| \ll |h_{33} h_{23}|$, so that the NP contribution to $B \to
K^{(*)} \nu {\bar\nu}$ is dominated by $ b \to s \nu_\tau
{\bar\nu}_\tau$. The constraint from $B \to K^{(*)} \nu {\bar\nu}$
then leads to $-0.047 \le (g_1 - g_3) \, h_{33} h_{23} \le 0.026$. For
the $S_3$ LQ, we have $h_{33} h_{23} = -0.28 \pm 0.08$ ($R_{D^{(*)}}$)
and $h_{33} h_{23} \ge -0.094$ ($B \to K^{(*)} \nu
{\bar\nu}$). Similarly, the $U_3$ LQ has $h_{33} h_{23} = -0.14 \pm
0.04$ ($R_{D^{(*)}}$) and $h_{33} h_{23} \ge -0.013$ ($B \to K^{(*)}
\nu {\bar\nu}$). In both cases, the two constraints on $h_{33} h_{23}$
are incompatible, so that the $S_3$ and $U_3$ LQ models cannot explain
the $B$-decay data.

Previous analyses \cite{RKRDmodels,Zurich} ruled out the $S_3$ and
$U_3$ models as candidate for explaining all the $B$-decay anomalies.
In these papers it was assumed that the NP couples predominantly to
the third generation. We have shown that the elimination of these
models is completely general -- even if the NP couplings involving the
second generation are allowed to be sizeable, the $S_3$ and $U_3$ LQ
models still cannot explain the $\bsmumu$ and $\bctaunu$ anomalies.

\subsection{$U_1$ LQ} 

\subsubsection{Fit}
\label{U1LQfit}

For the $U_1$ LQ, we perform a fit to all the $2q2\ell$ observables in
Table \ref{tab:obs_meas}, again taking $M_{\rm LQ} = 1$ TeV. There are
two important details. First, for $\tau \to \mu \phi$ there is only a
90\% C.L.\ upper limit on its branching ratio of $8.4 \times
10^{-8}$. In order to incorporate this observable into the fit, we
take $\cB(\tau \to \mu \phi) = (0.0 \pm 5.6) \times 10^{-8} $. Second,
note that the contribution to $\bsnunubar$ vanishes if $g_1 = g_3$
[see $C_L^{ij}({\rm NP})$ in Eq.~(\ref{NPWCs})].  But this is
precisely the definition of the $U_1$ model [Eq.~(\ref{g1g2})], so
there are no constraints on the $U_1$ LQ from this process. This
avoids the problem that eliminated the $S_3$ and $U_3$ LQ
models. Similarly, the $U_1$ LQ does not contribute to $J/\psi \to
\mu^\pm \tau^\mp$. There are thus 9 observables in the fit. As before,
the theoretical parameters are $h_{22}$, $h_{23}$, $h_{32}$ and
$h_{33}$, so that the d.o.f.\ is 5.

We find $\chi^2_{min,\SM + U_1} = 5.0$, or $\chi^2_{min}/{\rm d.o.f.}
= 1.0$.  This is an acceptable fit, so we see that the $U_1$ LQ model
does provide an explanation of the $B$-decay anomalies.

Now, the observables depend almost exclusively on products of the
couplings:
\bea
\label{2q2lprocesses}
\bsmumu &:& h_{32} h_{22} ~, \nn\\
\bctaunu &:& V_{cs} h_{33} h_{23} + V_{cb} h_{33}^2 ~, \nn\\
B^+ \to K^+ \tau^- \mu^+ &:& h_{32} h_{23} ~, \nn\\
B^+ \to K^+ \tau^+ \mu^- &:& h_{33} h_{22} ~, \nn\\
\Upsilon(2S) \to \mu^\pm \tau^\mp &:& h_{33} h_{32} ~, \nn\\
\tau \to \mu \phi &:& h_{23} h_{22} ~.
\eea
The only term that depends on a single coupling is the $h_{33}^2$
contribution in $\bctaunu$. But since it is multiplied by the small
CKM matrix element $V_{cb}$, its effect is small (unless $h_{33}$ is
quite large). And because only products of couplings are involved,
there is little information about the individual couplings themselves.

\begin{figure}[h]
\begin{center}
\includegraphics[width=0.4\textwidth]{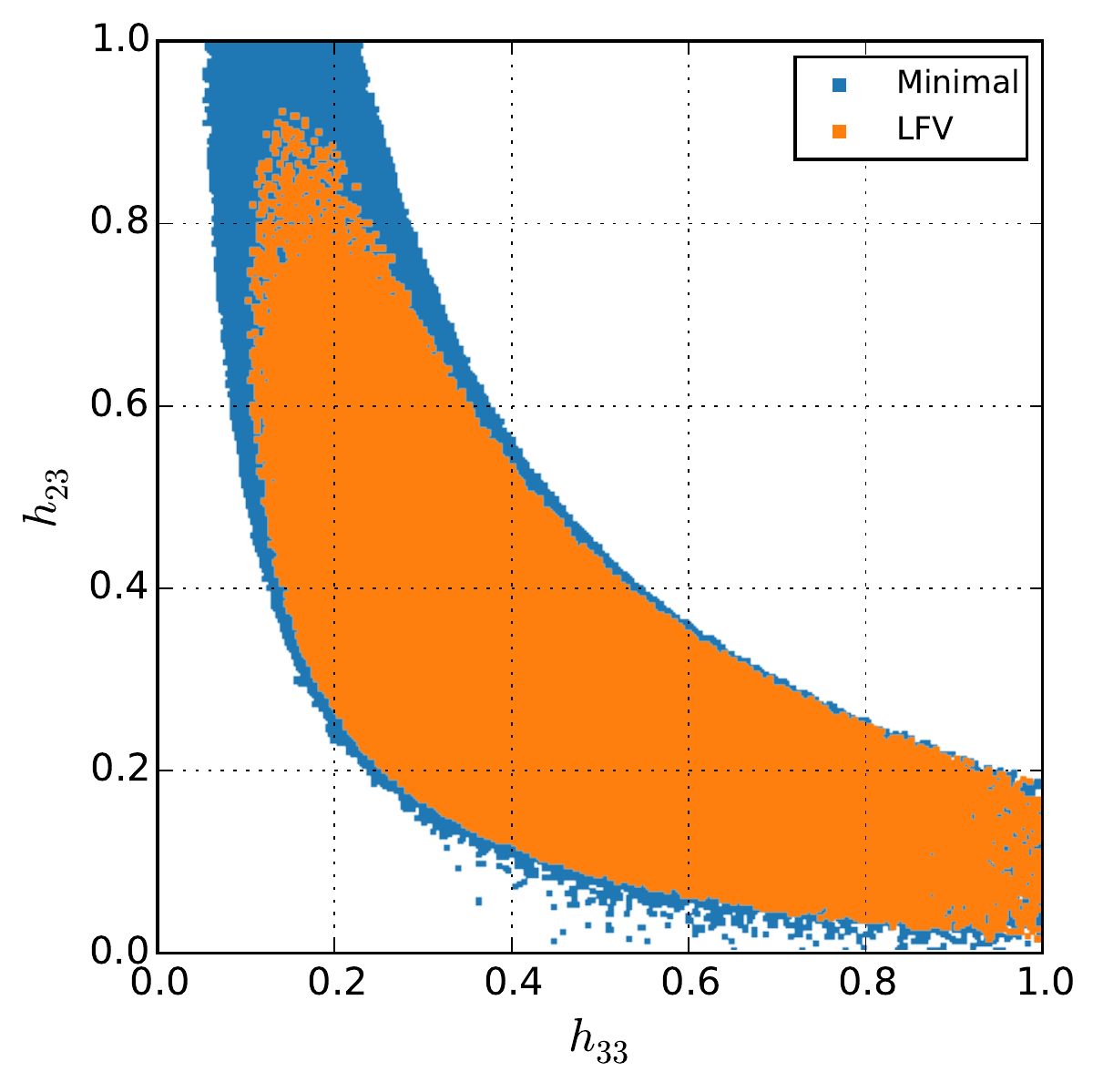} ~~~~~~
\includegraphics[width=0.4\textwidth]{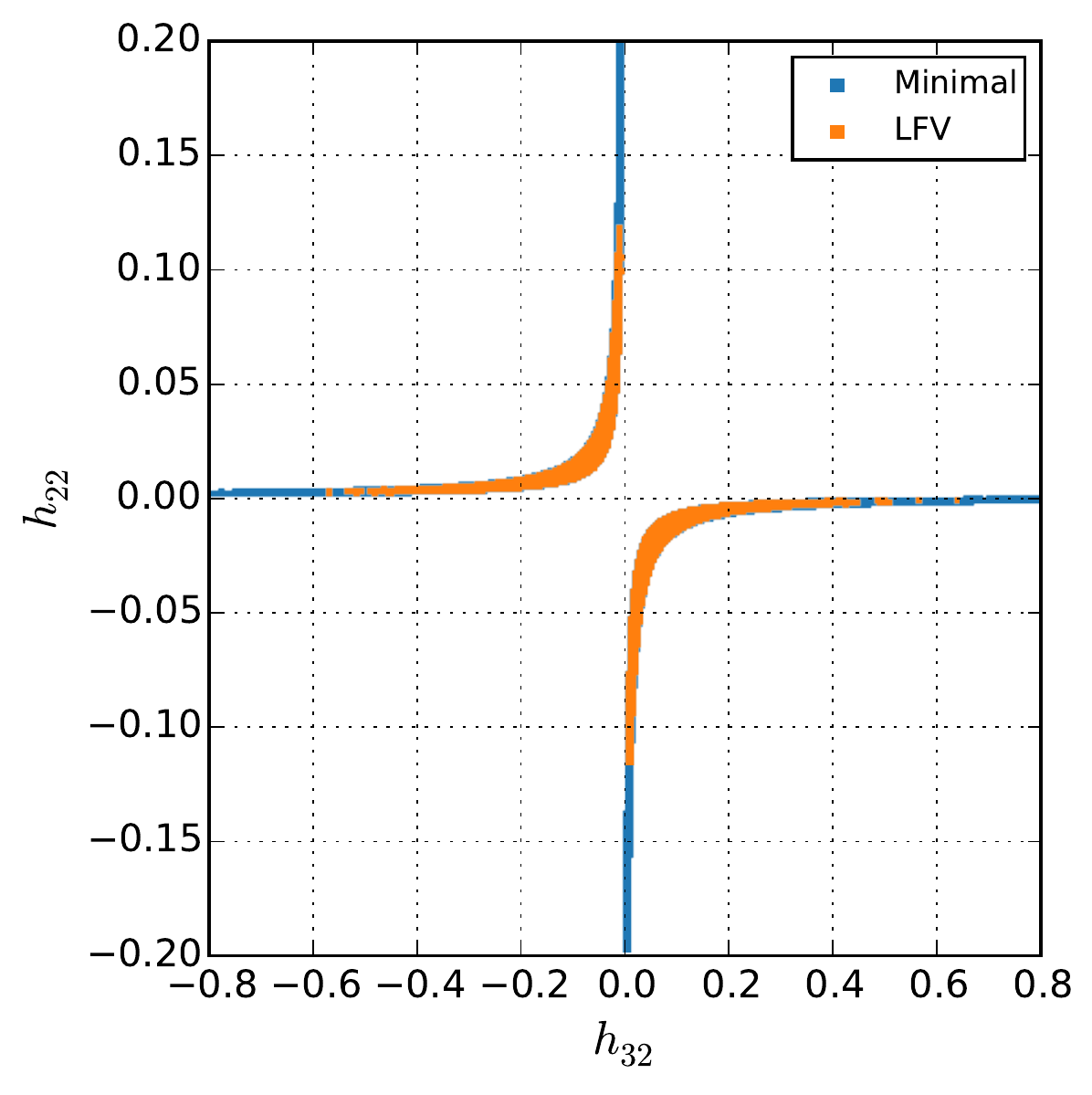}
\caption{Allowed 95\% C.L. regions in $h_{33}$-$h_{23}$ space (left
  plot) and $h_{32}$-$h_{22}$ space (right plot), for $M_{\rm LQ} = 1$
  TeV.  The regions are shown for a fit with only minimal constraints
  (blue) or minimal $+$ LFV constraints (orange).}
\label{fig:U1fit}
\end{center}
\end{figure}

This is illustrated in Fig.~\ref{fig:U1fit}, where we show the allowed
95\% C.L. regions in $h_{33}$-$h_{23}$ space (left plot)\footnote{In
  order for the $h_{ij}$ to be perturbative, we must have
  $h_{ij}^2/4\pi < 1$. To ensure this, we take the maximal value of
  the couplings to be $|h_{ij}| = 1$.} and in $h_{32}$-$h_{22}$ space
(right plot). These regions are determined largely by the $\bctaunu$
and $\bsmumu$ data, respectively. When one adds the LFV constraints,
the allowed regions are reduced in size, but are still sizeable.  The
LFV constraints place maximal values on some of the couplings:
$|h_{22}| \le 0.12$, $|h_{32}| \le 0.7$, and $h_{23} \le 0.9$. They
also lead to $h_{33} \ge 0.1$.

Some additional information can be learned by performing fits with
fixed values of $h_{33}$. In Table \ref{tab:fit_fixedh33}, we present
$\chi^2_{min,\SM + U_1}$ and the best-fit value of $h_{23}$ for
various values of $h_{33}$. We see that, as $h_{33}$ decreases and
$h_{23}$ increases, $\chi^2_{min,\SM + U_1}$ increases. This indicates
that the data prefer larger values of $h_{33}$ and smaller values of
$h_{23}$.

\begin{table*}[t] 
\begin{center}
\begin{tabular}{|c||c|c|}
 \hline\hline 
$h_{33}$ & $\chi^2_{min,SM + U_1}$ & $ h_{23}$ \\
\hline
$1.0$ & $5.0$ & $0.10 \pm 0.04$ \\
$0.5$ & $5.2$ & $0.26 \pm 0.07$ \\
$0.2$ & $6.8$ & $0.60 \pm 0.15$ \\
$0.1$ & $11.3$ & $0.70 \pm 0.20$ \\
 \hline\hline 
\end{tabular}
\end{center}
\caption{$U_1$ LQ model: $\chi^2_{min,\SM + U_1}$ and the best-fit
  value of $h_{23}$ for various values of $h_{33}$, for $M_{\rm LQ} =
  1$ TeV.}
\label{tab:fit_fixedh33}
\end{table*}

But this all raises a question. In the fit, we have seen that the LFV
constraints put maximal values on some of the couplings. Is this the
only effect of the LFV observables?  The answer is no. Because the LFV
processes of Eq.~(\ref{2q2lprocesses}) involve one of $\{
h_{33},h_{23} \}$ and one of $\{ h_{32},h_{22} \}$, they relate
portions of the $h_{33}$-$h_{23}$ and $h_{32}$-$h_{22}$ regions. And,
in fact, these relations can be quite important.

\begin{figure}[h]
\begin{center}
\includegraphics[width=0.4\textwidth]{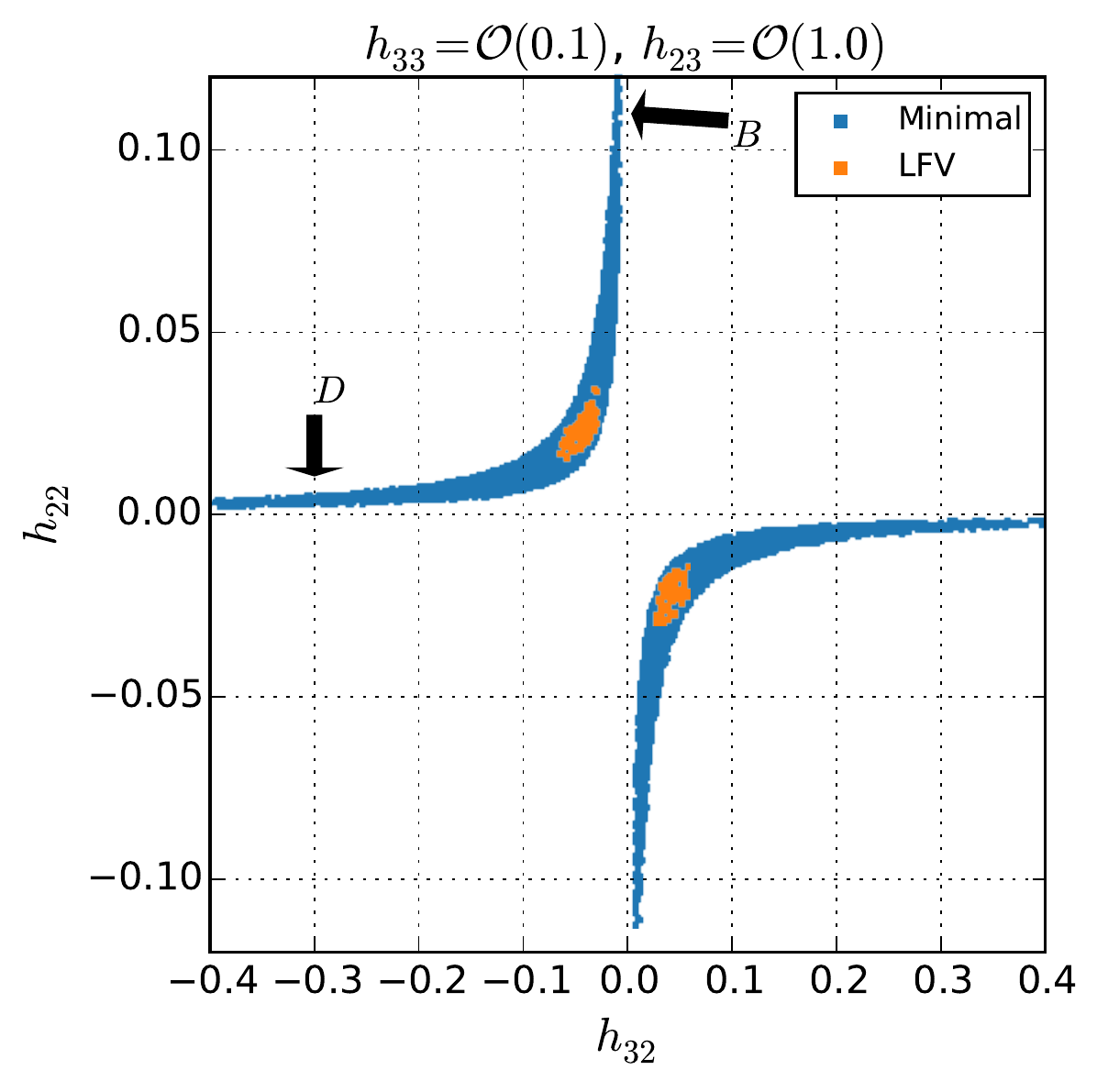} ~~~~~~
\includegraphics[width=0.4\textwidth]{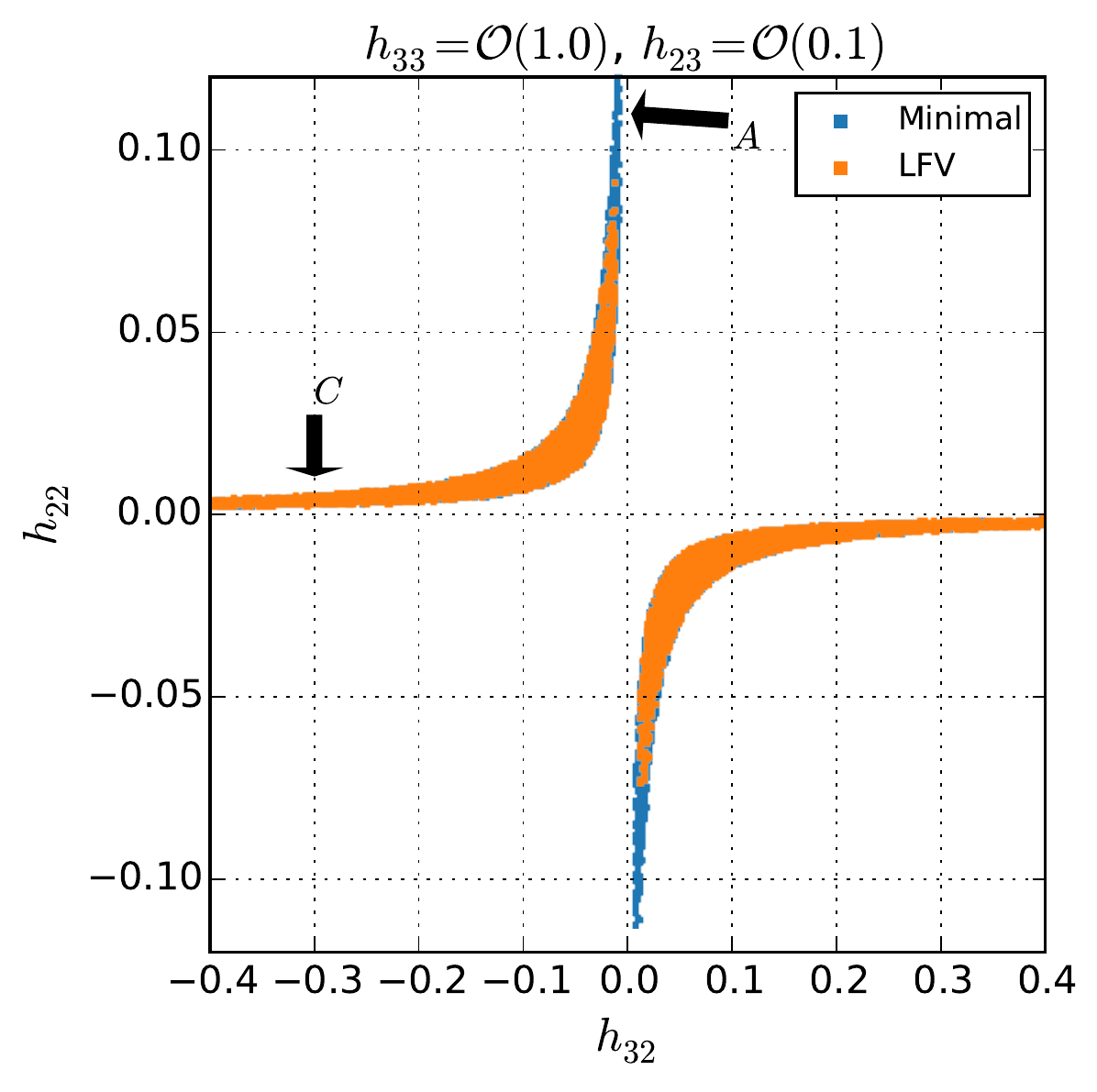}
\caption{Allowed 95\% C.L. region in $h_{32}$-$h_{22}$ space for $\{
  h_{33},h_{23} \} = \{ O(0.1), O(1.0) \}$ (left plot) or $\{
  h_{33},h_{23} \} = \{ O(1.0), O(0.1) \}$ (right plot), for $M_{\rm
    LQ} = 1$ TeV. The region is shown for a fit with only minimal
  constraints (blue) or minimal $+$ LFV constraints (orange).}
\label{fig:ABCD}
\end{center}
\end{figure}

To illustrate this, we note that the $\bctaunu$ data imply $h_{33}
h_{23} = 0.14 \pm 0.04 = O(0.1)$ for $M_{\rm LQ} = 1$ TeV. To
reproduce this, we consider two limiting cases: $\{ h_{33},h_{23} \}
=$ (a) $\{ 0.1, 1.0 \}$ or (b) $\{ 1.0, 0.1 \}$. Also, the $\bsmumu$
data lead to $h_{32} h_{22} = -0.0011 \pm 0.0002 = O(0.001)$. In the
same vein, we consider two limiting cases: $\{ h_{32},h_{22} \} =$ (c)
$\{ O(0.01), O(0.1) \}$ or (d) $\{ O(0.1), O(0.01) \}$. These can be
combined to produce four rough scenarios for the four couplings:
\bea
A = (a,c) &:& h_{33} = O(1.0) ~,~~ h_{23} = O(0.1) ~,~~ h_{32} = O(0.01) ~,~~ h_{22} = O(0.1) ~, \nn\\ 
B = (b,c) &:& h_{33} = O(0.1) ~,~~ h_{23} = O(1.0) ~,~~ h_{32} = O(0.01) ~,~~ h_{22} = O(0.1) ~, \nn\\ 
C = (a,d) &:& h_{33} = O(1.0) ~,~~ h_{23} = O(0.1) ~,~~ h_{32} = O(0.1) ~,~~ h_{22} = O(0.01) ~, \nn\\ 
D = (b,d) &:& h_{33} = O(0.1) ~,~~ h_{23} = O(1.0) ~,~~ h_{32} = O(0.1) ~,~~ h_{22} = O(0.01) ~.
\eea
We now repeat the fit, fixing the couplings $h_{33}$ and $h_{23}$ as
per (a) or (b). In addition, the fit is performed using (i) only the
minimal constraints or (ii) the minimal $+$ LFV constraints. The
allowed 95\% C.L. regions in $h_{32}$-$h_{22}$ space are shown in
Fig.~\ref{fig:ABCD}, with case (a) on the left and case (b) on the
right. If only minimal constraints are used, there is no difference
between (a) and (b) -- the allowed region is the same in both cases,
and scenarios $A$, $B$, $C$ and $D$ are all allowed. However, this
changes when the LFV constraints are added. For case (a), the allowed
region is greatly reduced: $h_{32}$ and $h_{22}$ must both be rather
small, and scenarios $B$ and $D$ are both ruled out. On the other
hand, the effect of the addition of the LFV constraints is much less
dramatic for case (b). Most of the region allowed with minimal
constraints is still allowed, though scenario $A$ is now ruled
out. This demonstrates the effect that the LFV constraints have on the
parameter space.

We have emphasized that previous analyses made the theoretical
assumption that the NP couples predominantly to the third generation.
This implies a large value of $h_{33}$. Now, above we noted that the
data prefer larger values of $h_{33}$. This suggests that, in fact,
such a theoretical assumption is not necessary -- the data point in
this direction. How does this come about?  After all, the $\bctaunu$
constraints depend essentially on the product $h_{33} h_{23}$
[Eq.~(\ref{2q2lprocesses})]. So a small value of $h_{33}$ can be
compensated for by a large value of $h_{23}$. However, we saw above
that such a scenario is disfavoured by the LFV constraints. Indeed, it
is these LFV constraints that lead to the requirement of a large value
of $h_{33}$, in line with the theoretical assumption.

\subsubsection{Renormalization group equations} 
\label{RGEanalysis}

In Refs.~\cite{Feruglio:2016gvd,Feruglio:2017rjo}, additional
constraints were derived. The starting point is the observation that
the scale of NP, $\Lambda$, is well above the weak scale $v$ [e.g.,
$\Lambda =$ O(TeV)]. Below $\Lambda$, but above $v$, the physics is
described by ${\cal L}_\SM + {\cal L}_\NP$. Here ${\cal L}_\NP$ is the
effective Lagrangian obtained when the NP is integrated out; it is
invariant under the SM gauge group. In
Refs.~\cite{Feruglio:2016gvd,Feruglio:2017rjo}, it was assumed that
the dominant terms in ${\cal L}_\NP$ are the $2q2\ell$ operators of
Eq.~(\ref{2q2lops}), written in the weak basis, with the NP coupling
only to the third generation. Once $SU(2)_L \times U(1)_Y$ is broken,
the fermions acquire masses. One transforms from the weak basis to the
mass basis by acting on the fermion fields with unitary
transformations. In the mass basis, the NP couplings are functions of
these transformations and the couplings in the weak basis.

${\cal L}_\NP$ is evolved from the NP scale $\Lambda$ to the weak
scale using the one-loop renormalization group equations (RGEs) in the
limit of exact electroweak symmetry. After performing a matching at
the weak scale, it is further evolved down to the scale of 1 GeV using
the QED RGEs and integrating out the heavy degrees of freedom.

This evolution has several effects. First, for the $U_1$ LQ model,
recall that the constraints from $\bsnunubar$ could be evaded because
$g_1 = g_3$. However, this equality holds only at the NP scale
$\Lambda$. At lower energies, a nonzero value of $\delta g_- \equiv
g_1 - g_3$ is generated. This means that constraints from $B \to
K^{(*)} \nu {\bar\nu}$ must be taken into account for $U_1$.

Second, there is operator mixing during the RGE evolution. One of the
effects is that the leptonic couplings of the $W^\pm$ and $Z^0$ are
modified. This can be understood as follows. If one combines the SM
decay $Z^0 \to q {\bar q}$ with the NP process $q {\bar q} \to \ell_i
{\bar\ell}_j$, this corresponds to a (loop-level) NP contribution to
$Z^0 \to \ell_i {\bar\ell}_j$. If $i = j$, this is a correction to the
coupling of the $Z^0$ to charged or neutral leptons. And if $i \ne j$,
this generates an LFV decay of the $Z^0$. There are similar effects
for the coupling of the $Z^0$ to quarks, and all this also holds for
the $W^\pm$. However, since the leptonic couplings of the $Z^0$ are
the most precisely measured, the constraints from these measurements
are the most important.

Another effect of this operator mixing is that, at low energies, when
the $W$, $Z$, $t$, $b$ and $c$ have all been integrated out, one
generates four-fermion LFV processes such as $\tau \to 3 \mu$, $\tau
\to \mu \rho$ and $\tau \to \mu \pi$, as well as corrections to the
LFC decay $\tau \to \ell \nu_\tau {\bar\nu}_\ell$. In the case of
$\tau \to 3 \mu$, this can be understood as the combination of SM $q
{\bar q} \mu^+ \mu^-$ and NP $q {\bar q} \mu \tau$ operators.

Two scenarios are examined in Ref.~\cite{Feruglio:2016gvd}: (i) $g_1 =
0$ and $|g_3| \le 3$, (ii) $g_1 = g_2$. It is argued that the new RGE
constraints are very important, particularly for scenario (i). In
Ref.~\cite{Feruglio:2017rjo}, under the additional assumptions that
the mass-basis couplings obey $h_{33} = 1$, $h_{23} = h_{32}$ and
$h_{22} = h_{23}^2$, it was shown that the RGE constraints rule out
scenario (ii) entirely, mostly due to the constraints from $\tau \to
\ell \nu {\bar\nu}$. (We note that the assumptions about the couplings
correspond to an extremely special case, where the transformations
from the weak to the mass basis are the same for down-type quarks and
charged leptons.)

We have several observations regarding the above RGE analysis:
\begin{itemize}

\item The analysis of Refs.~\cite{Feruglio:2016gvd,Feruglio:2017rjo}
  is at the level of an effective field theory (EFT). As such, the
  results of this analysis are not necessarily applicable to all
  models, since a given model may have additional operators in ${\cal
    L}_\NP$. These extra operators may affect the RGEs and the
  conclusions.

\item As a specific example, the $VB$ model has $g_1 = 0$, and so one
  might think it is represented by scenario (i) above. This is not
  true: the $VB$ model also has tree-level four-quark and four-lepton
  operators. In particular, there is a tree-level contribution to
  $\tau \to 3\mu$. In this case, the RGE generation of a (loop-level)
  contribution to $\tau \to 3\mu$ is irrelevant.

\item A similar comment applies to the EFT analysis itself. Much
  emphasis is placed on the RGE generation of contributions to LFV
  processes such as $\tau \to 3\mu$, $\tau \to \mu \rho$, etc.
  However, all of these processes arise due to the combination of a SM
  operator with the NP operator $({\bar q}_{iL} \gamma_\mu q_{jL})
  ({\bar \mu}_L \gamma^\mu \tau_L)$. But the very existence of this NP
  operator leads to tree-level LFV effects in $B \to K \tau \mu$,
  $\tau \to \mu \phi$, $\Upsilon \to \mu \tau$ and $J/\psi \to \mu
  \tau$. There are stringent upper bounds on the branching ratios of
  all of these processes (see Table \ref{tab:obs_meas}). The upshot is
  that there is no need to consider the loop-level RGE effects -- the
  constraints on the NP operator coming from these tree-level
  processes are stronger.

\item Finally, the EFT analysis also leads to NP contributions to LFC
  processes such as $Z^0 \to \ell^+ \ell^-$, $Z^0 \to \nu_\ell
  {\bar\nu}_\ell$ and $\tau \to \ell \nu_\tau {\bar\nu}_\ell$. These
  processes are all measured quite precisely, so that, even though the
  NP contributions are small (loop level), they can be constrained by
  the measurements. While this conclusion is valid for the EFT, it
  does not necessarily hold in a real model. Consider the $U_1$ LQ. It
  contributes at one loop to all of these processes, so that, once the
  NP is integrated out, there are new operators in ${\cal L}_\NP$.
  Compared to the $2q2\ell$ operators of Eq.~(\ref{2q2lops}), they are
  indeed subdominant. However, they are of the same order as the
  low-energy RGE effects, so that there may be a partial cancellation
  between the two contributions. The bottom line is that the RGE
  constraints from LFC processes must be taken with a grain of salt --
  they may be evaded in real models. (To be fair, this is acknowledged
  explicitly in Ref.~\cite{Feruglio:2017rjo}.)

\end{itemize}
Our conclusion is that, while the RGE analysis of
Refs.~\cite{Feruglio:2016gvd,Feruglio:2017rjo} is interesting, the
results are suspect because the tree-level LFV constraints have not
been properly taken into account. And even if they are, one has to be
very careful about taking its constraints too literally, as they may
not hold in real models.

This said, in order to compare with previous analyses, we apply the
RGE analysis to our $U_1$ model, taking $M_{\rm LQ} = 1$ TeV. In our
general study, (i) we do not assume that the NP couples only to the
third generation in the weak basis, and (ii) we work in the mass
basis. In order to repeat the RGE analysis, but with our setup, we use
the programs {\tt Wilson} \cite{Aebischer:2018bkb} and {\tt flavio}
\cite{flavio}. The RGE constraints arise from the contributions to LFV
$\tau$ decays, $Z$-pole observables and $\tau \to \ell \nu_\tau
     {\bar\nu}_\ell$ ($\ell = e, \mu$). (Note that we have verified
     that {\tt Wilson} and {\tt flavio} reproduce previous
     calculations of the RGE constraints \cite{Feruglio:2017rjo,
       Zurich}.)

\begin{figure}[h]
\begin{center}
\includegraphics[width=0.4\textwidth]{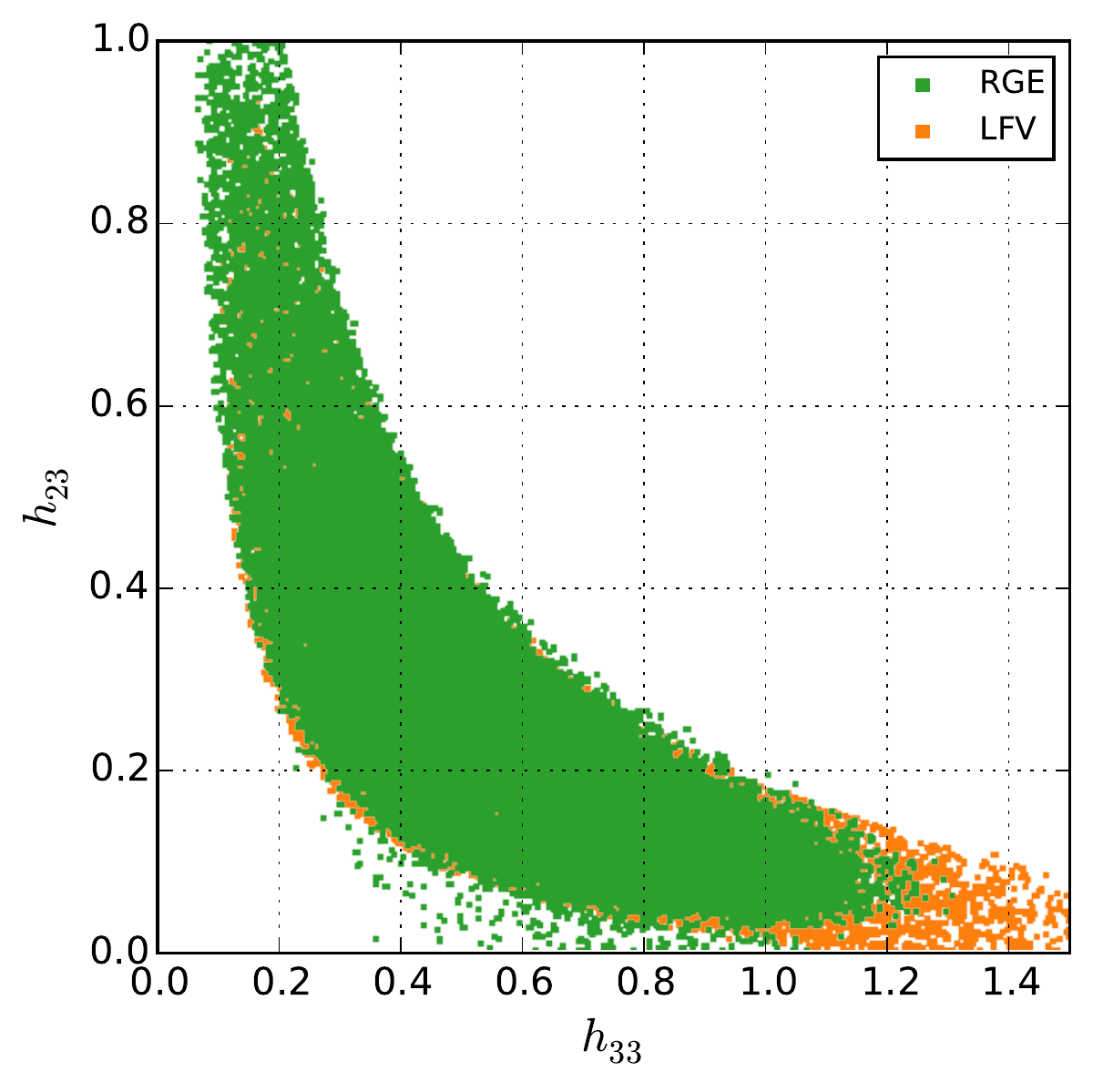} ~~~~~~
\includegraphics[width=0.4\textwidth]{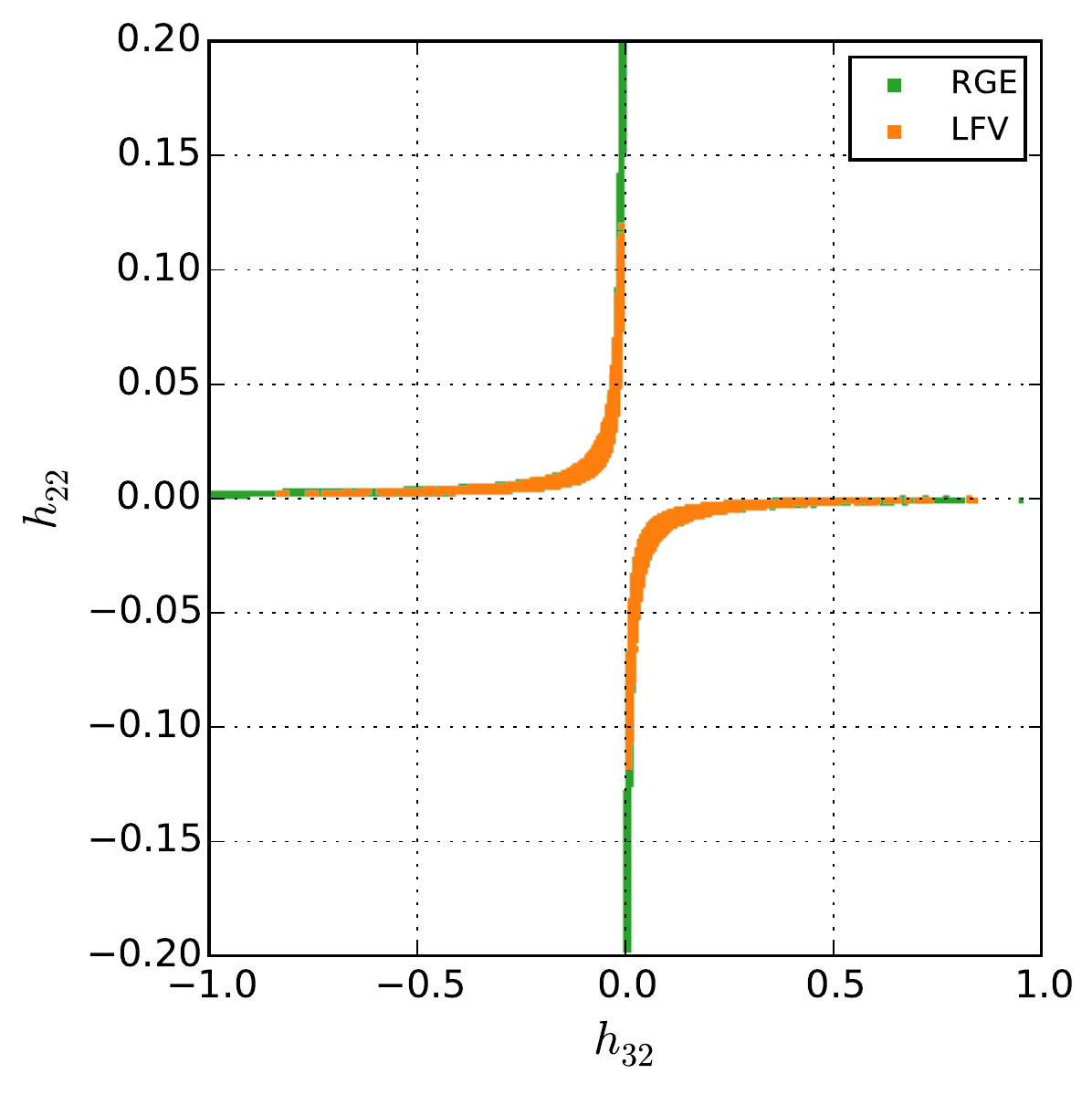}
\caption{Allowed 95\% C.L. regions in $h_{33}$-$h_{23}$ space (left
  plot) and $h_{32}$-$h_{22}$ space (right plot), for $M_{\rm LQ} = 1$
  TeV.  The regions are shown for a fit with minimal $+$ LFV
  constraints (orange) or minimal $+$ RGE constraints (green).}
\label{fig:U1fitRGE}
\end{center}
\end{figure}

In Fig.~\ref{fig:U1fitRGE}, we show the allowed 95\% C.L. regions in
$h_{33}$-$h_{23}$ space (left plot) and in $h_{32}$-$h_{22}$ space
(right plot), when the RGE (green) or LFV (orange) constraints are
added to the minimal constraints\footnote{We note that, if one
  compares Fig.~\ref{fig:U1fitRGE} with the equivalent figure in
  Ref.~\cite{Zurich}, the regions with RGE constraints don't look the
  same. However, this is because different notations are used. We vary
  the couplings $h_{ij}$ ($ij = 22, 23, 32, 33$), while the couplings
  in Ref.~\cite{Zurich} are $g_U \beta_{ij}$ ($ij = 22, 23, 32, 33$),
  with $\beta_{33}$ fixed to 1 and the coupling constant $g_U$ allowed
  to vary. If one takes into account this change of notation, it is
  found that the region with RGE constraints is very similar in the
  two analyses.}. One sees from these plots that, in general, the LFV
constraints are more stringent than the RGE constraints.  For example,
the LFV constraints lead to $|h_{22}| \le 0.12$, $|h_{32}| \le 0.7$,
and $h_{23} \le 0.9$, whereas the RGE constraints allow all of these
couplings to be as large as 1. Also, one has $h_{33} \ge 0.1$ with the
LFV constraints, while the RGE constraints allow this coupling to be
slightly smaller. The only coupling value for which this behaviour
does not hold is the maximal value of $h_{33}$. The RGE constraints
require $h_{33} \le 1.3$, while the LFV constraints allow much larger
values. This said, such large couplings are entering the
nonperturbative regime, which is why we previously imposed an upper
limit of 1 on the absolute value of all couplings. Thus, if one
requires $|h_{ij}| \le 1$, the RGE constraints are irrelevant compared
to the LFV constraints.

\subsubsection{$\bs$-$\bsbar$ mixing}
\label{LQBsmix}

In Sec.~\ref{RGEanalysis} above, we saw that the $U_1$ LQ can
contribute at one loop to four-lepton operators, and there can
potentially be some constraints from the measurements of such
processes. In the same vein, there can also be one-loop contributions
to four-quark processes. From the point of view of constraining the
$U_1$ model, the most promising four-quark observable is
$\bs$-$\bsbar$ mixing. Does its measurement, characterized by $\Delta
M_s$, yield constraints on the $U_1$ LQ?

In the SM, the underlying quark-level process can be accurately
computed. However, there is a hadronic uncertainty in converting this
to the level of mesons. This is described in detail in
Sec.~\ref{Addobs}, but here we summarize the main points.  The
relevant hadronic parameter is $f_{B_s}\sqrt{\hat B_{B_s}} = (266 \pm
18)$ MeV \cite{Aoki:2016frl}. The central value is such that the SM
reproduces the measured value of $\Delta M_s$. However, the error is
sufficiently large that there is some room for NP. As a consequence, a
small, loop-level NP contribution is allowed. That is, there are no
constraints on the $U_1$ LQ model from $\bs$-$\bsbar$ mixing.

Recently, the hadronic parameters were recalculated, and larger values
were found \cite{Bazavov:2016nty}. The implications for $\bs$-$\bsbar$
mixing were examined in Ref.~\cite{DiLuzio:2017fdq}. It was found that
the central value of the SM prediction for $\Delta M_s$ is now
$1.8\sigma$ above its measured value. This led the authors of
Ref.~\cite{DiLuzio:2017fdq} to observe that this poses problems for NP
solutions of the $\bsmumu$ anomalies. The point is that, regardless of
whether the NP is a $Z'$ or a LQ, the contribution to $\Delta M_s$ has
the same sign as that of the SM. That is, the discrepancy with
measurement increases in the presence of NP. The problem is
particularly severe for the $Z'$, where the contribution to
$\bs$-$\bsbar$ mixing is tree level, and hence large. But it also
applies to the LQ, whose contribution is loop level.

If this new calculation is correct, it does indeed create problems for
NP solutions of the $B$ anomalies. But it also creates important
problems for the SM. The SU(3)-breaking ratio of hadronic matrix
elements in the $\bs$ and $\bd$ systems is well known: $\xi = 1.206
\pm 0.018 \pm 0.006$ \cite{Bazavov:2016nty}. If the SM prediction of
$\Delta M_s$ is in disagreement with its measured value, the same
holds for $\Delta M_d$. And this has important consequences for fits
to the CKM matrix \cite{CKMfitter}.

Thus, the results of the new calculation of the hadronic parameters
may have important implications for the SM. In light of this, we
prefer to wait for a verification of the new result before including
it among the constraints on the $U_1$ LQ model.

\subsection{Predictions}
\label{Sec:LQPredictions}

Having established that the $U_1$ LQ model can explain the $B$
anomalies, the next step is to examine ways of testing this
explanation. To this end, here we present the predictions of the
model.

\begin{figure}[h]
\begin{center}
\includegraphics[width=0.4\textwidth]{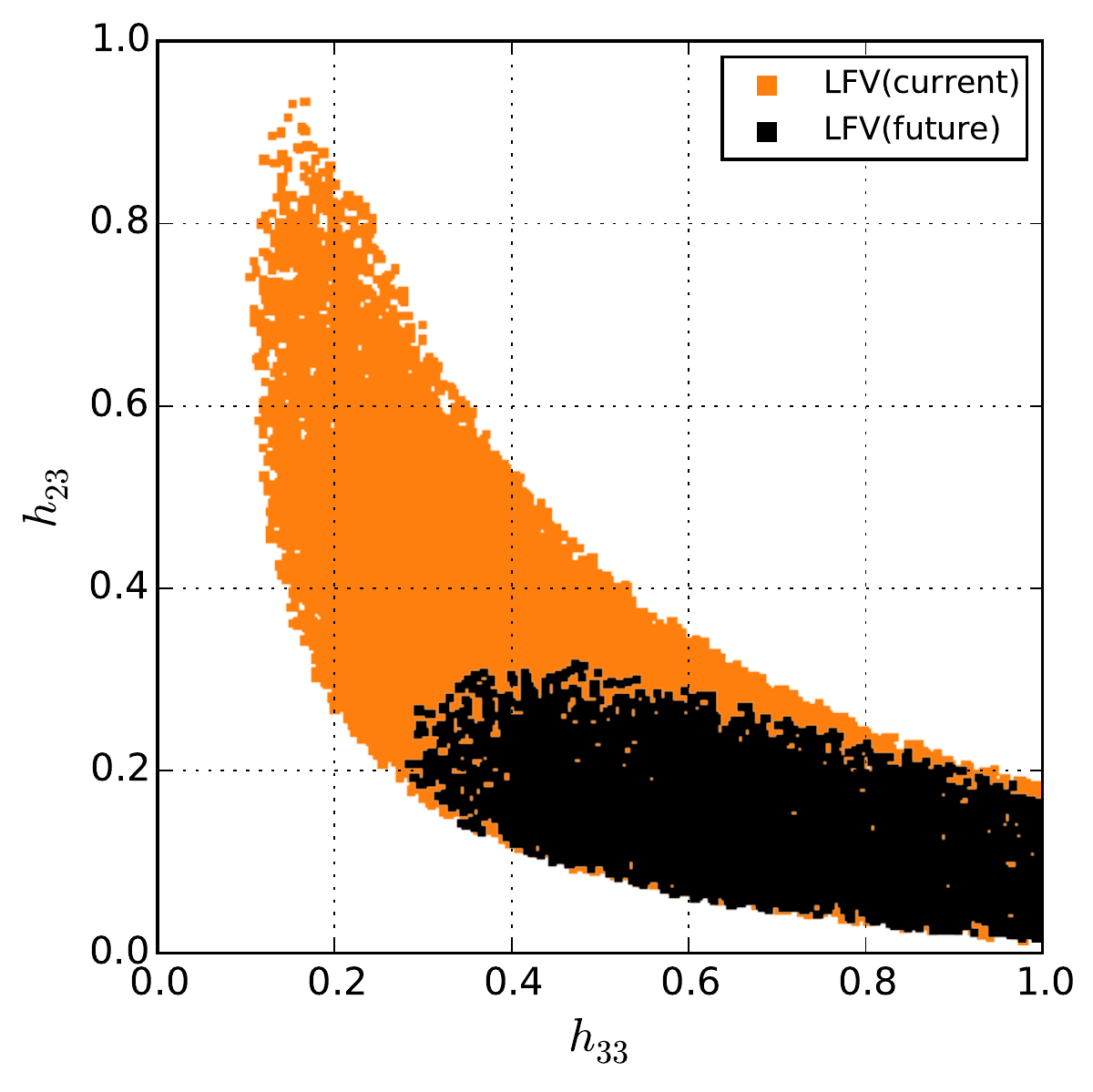} ~~~~~~
\includegraphics[width=0.4\textwidth]{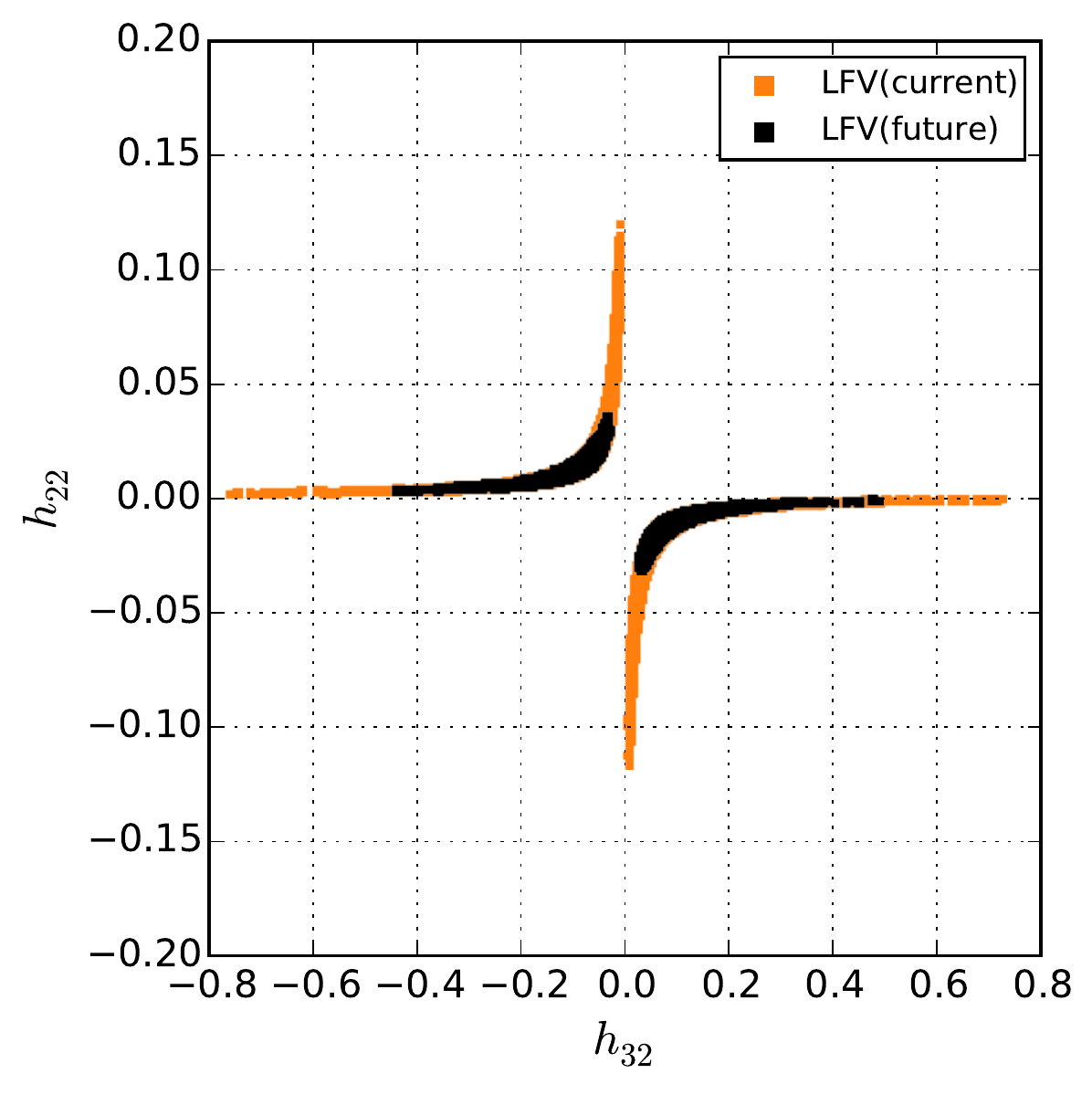}
\caption{Allowed 95\% C.L. regions in $h_{33}$-$h_{23}$ space (left
  plot) and $h_{32}$-$h_{22}$ space (right plot), for $M_{\rm LQ} = 1$
  TeV.  The regions are shown for a fit with the minimal constraints
  $+$ present LFV constraints (orange) or future LFV constraints
  (black).}
\label{fig:U1fitp}
\end{center}
\end{figure}

Above, we have emphasized the importance of the semileptonic LFV
processes $B \to K^{(*)} \mu^\pm \tau^\mp$, $\tau \to \mu \phi$ and
$\Upsilon \to \mu\tau$. To date, no such decay has been
observed. However, this may change in the future. For example, the
expected reach of Belle II is $\cB(B^+ \to K^+ \mu^\pm \tau^\mp) = 3.3
\times 10^{-6}$, ${\cal B}(\Upsilon \to \mu^\pm \tau^\mp) = 1.0 \times
10^{-7}$ and ${\cal B}(\tau \to \mu \phi) = 1.5 \times 10^{-9}$
\cite{BelleIIreach}.  Does the $U_1$ model predict that at least one
of these decays will be observed at Belle II? Unfortunately, the
answer is no. In Fig.~\ref{fig:U1fitp}, we show the allowed 95\%
C.L. regions in $h_{33}$-$h_{23}$ space (left plot) and in
$h_{32}$-$h_{22}$ space (right plot) for the case where {\it no} LFV
signal is observed, i.e., where the above reaches are applied as upper
limits. As can be seen from the figures, although the allowed space of
couplings would be reduced, it is still sizeable.  That is, if the
$U_1$ LQ model is the correct explanation of the $B$ anomalies, an LFV
process may be observed at Belle II, but there is no guarantee.

Other observables are more promising. The measurement of $R_{D^{(*)}}$
corresponds to LFUV in $b \to c \ell^- {\bar\nu}_\ell$. The NP effect
is mainly for $\ell=\tau$ and is governed by $V_{cs} h_{33} h_{23} +
V_{cb} h_{33}^2$ [Eq.~(\ref{2q2lprocesses})]. One then also expects to
observe LFUV in $b \to u \ell^- {\bar\nu}_\ell$, with the NP
contribution proportional to $V_{us} h_{33} h_{23} + V_{ub} h_{33}^2$.
Such an effect can be seen in $B \to \pi \ell {\bar\nu}_\ell$ or $B^-
\to \ell {\bar\nu}_\ell$ decays \cite{Tanaka:2016ijq}. The observables
are denoted $R_{\pi\ell{\bar\nu}}^{\tau/\mu}$ and
$R_{\ell{\bar\nu}}^{\tau/\mu}$, respectively. Another process where
one expects significant NP effects is $b \to s \tau^+ \tau^-$. Here
the decays are $B \to K^{(*)} \tau^+ \tau^-$ and $\bs \to \tau^+
\tau^-$.  Finally, there is $B \to K^{(*)} \nu {\bar\nu}$, whose
fermion-level decay is $\bsnunubar$. As discussed in
Sec.~\ref{RGEanalysis}, at low-energies there is a contribution to
this decay from the $U_1$ LQ, due to the evolution of the RGEs. As
noted in Sec.~\ref{RGEanalysis}, one must take this calculation with a
grain of salt, since there may be additional contributing operators at
the NP scale.

\begin{figure}[h]
\begin{center}
\includegraphics[width=0.4\textwidth]{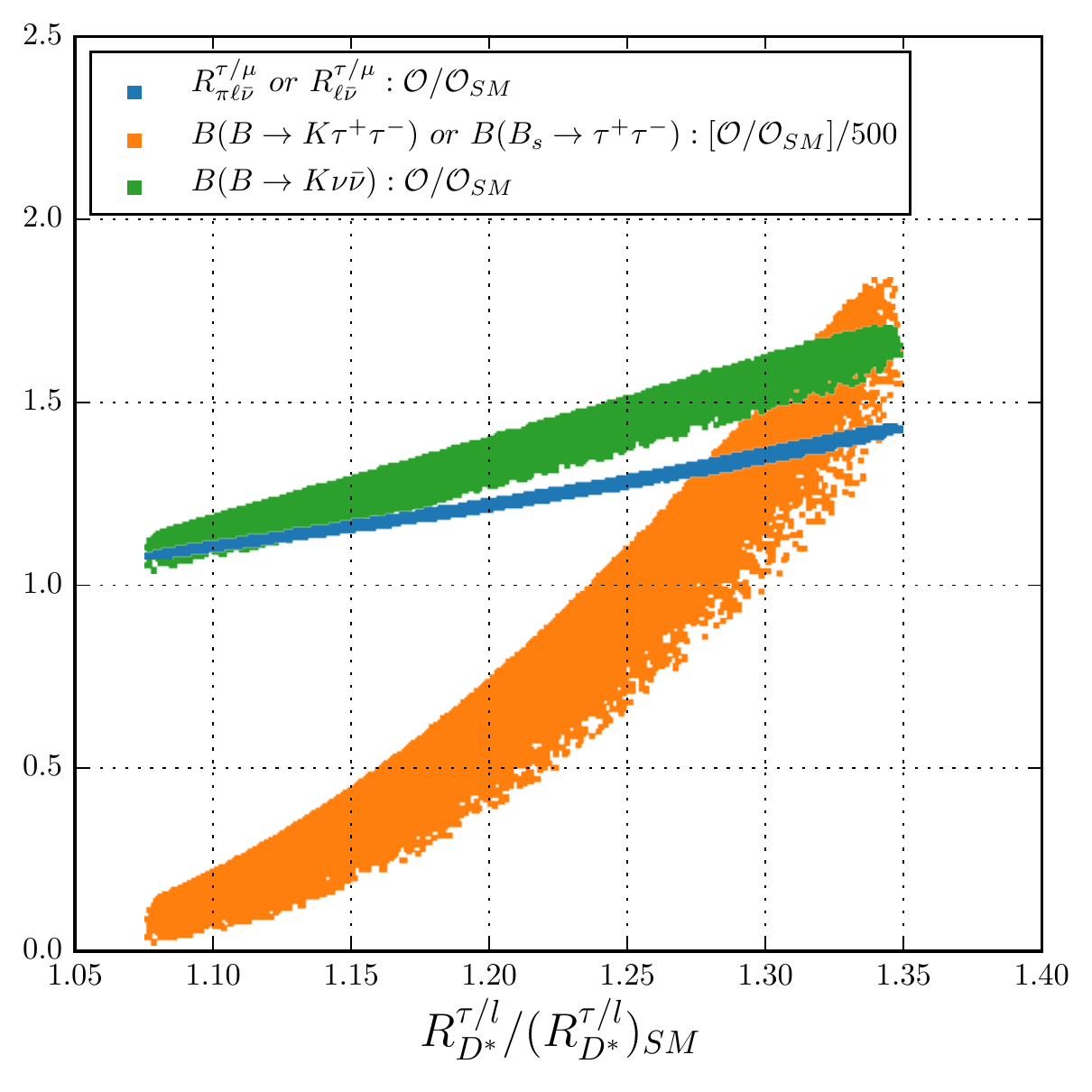}
\caption{Within the $U_1$ LQ model with $M_{\rm LQ} = 1$ TeV,
  predictions for observables as a function of the value of
  $R_{D^*}^{\tau/\ell}/(R_{D^*}^{\tau/\ell})_\SM$. Observables ${\cal
    O}$ are: $R_{\pi\ell{\bar\nu}}^{\tau/\mu}$ or
  $R_{\ell{\bar\nu}}^{\tau/\mu}$ (blue), $\cB(B \to K^{(*)} \tau^+
  \tau^-)$ or $\cB(\bs \to \tau^+ \tau^-)$ (orange), and $\cB(B \to
  K^{(*)} \nu {\bar\nu})$ (green). Quantities plotted are ${\cal
    O}/{\cal O}_\SM$ (blue and green) or $[{\cal O}/{\cal O}_\SM]/500$
  (orange).}
\label{LQpredictions}
\end{center}
\end{figure}

The predictions for all these observables are shown in
Fig.~\ref{LQpredictions} as a function of the value of
$R_{D^*}^{\tau/\ell}/(R_{D^*}^{\tau/\ell})_\SM$, for $M_{\rm LQ} = 1$
TeV. For all three observables, there may be a significant enhancement
compared to the SM predictions. $R_{\pi\ell{\bar\nu}}^{\tau/\mu}$ and
$R_{\ell{\bar\nu}}^{\tau/\mu}$ can be larger by as much as 40\%, while
$\cB(B \to K^{(*)} \nu {\bar\nu})$ may be increased by 70\% over the
SM. As for $\cB(B \to K^{(*)} \tau^+ \tau^-)$ and ${\cal B}(\bs \to
\tau^+ \tau^-)$, they can be enhanced by as much as three orders of
magnitude. This is consistent with the findings of
Refs.~\cite{AGC,Crivellin:2017zlb,Calibbi:2017qbu}. (Ref.~\cite{Capdevila:2017iqn}
discusses using $b \to s \tau^+ \tau^-$ to search for NP.)

One key feature of Fig.~\ref{LQpredictions} is that these predictions
are correlated with one another, and with the value of
$R_{D^*}^{\tau/\ell}/(R_{D^*}^{\tau/\ell})_\SM$. The reason is that
the NP contributions to all four observables are either dominated by
$h_{23} h_{33}$ ($\bstautau$, $\bsnunubar$) or have $h_{23} h_{33}$ as
the main component ($b \to c \ell^- {\bar\nu}_\ell$, $b \to u \ell^-
{\bar\nu}_\ell$). Now, $R_{D^*}^{\tau/\ell}$ will be remeasured with
greater precision. If the deviation of its value from the SM
prediction is found to be large (small), the deviations of the other
observables from their SM predictions are also predicted to be large
(small). This is a good test of the $U_1$ LQ model.

\section{Vector boson model}
\label{Sec:VBmodel}

This model contains SM-like vector bosons ($VB$s) that transform as
$({\bf 1},{\bf 3},0)$ under $SU(3)_C \times SU(2)_L \times U(1)_Y$.
The $VB$s are denoted $V = W', Z'$. In the mass basis, the Lagrangian
describing the couplings of the $VB$s to left-handed fermions is
\bea
\De\cL^{}_{V} &=& g^q_{ij}\(\oQ_{iL}~\ga^\mu\si^I~Q_{jL}\)V^{I}_{\mu}
~+~ g^\ell_{ij}\(\oL_{iL}~\ga^\mu\si^I~L_{jL}\)V^{I}_{\mu}~.
\eea
Integrating out the heavy $VB$s, we obtain the following effective
Lagrangian, relevant for $2q2\ell$ decays:
\beq
\cL^\eff_{V} = - \frac{g_{ij}^q g_{kl}^\ell}{M^2_{V}}\(\oQ_{iL}\ga^\mu
\si^I~Q_{jL}\)\(\oL_{kL}\ga^{}_\mu\si^I L_{lL}\) ~.
\eeq
Comparing this with Eq.~(\ref{2q2lops}), we find
\beq
G_1^{ijkl} = 0 ~~,~~~~ G_3^{ijkl} = - g_{ij}^q g_{kl}^\ell ~.
\eeq

As was done with the LQ models, we take the couplings
$g_{ij}^{q,\ell}$ to be real, and assume that the $VB$ couplings to
the first-generation leptons and down-type quarks are negligible. In
the quark sector, there are then three independent couplings:
$g_{ss}$, $g_{sb} = g_{bs}$ and $g_{bb}$.  Similarly, in the lepton
sector, the three independent couplings are $g_{\mu\mu}$, $g_{\mu\tau}
= g_{\tau\mu}$ and $g_{\tau\tau}$. For the leptons, these couplings
hold for either component of the $SU(2)_L$ doublet. Thus, for example,
$g_{\mu\mu} = g_{\nu_\mu \nu_\mu} = g_{\mu \nu_\mu}$. The quark sector
is a bit more complicated because the couplings to the up-type quarks
involve the CKM matrix [Eq.~(\ref{NPcouplings})]. For example, for the
$W'$, this implies $g_{cb} = V_{cs} \, g_{sb} + V_{cb} \, g_{bb}$,
while for the $Z'$, we have $g_{cc} = V_{cs}^2 \, g_{ss} + 2 V_{cs}
V_{cb} \, g_{sb} + V_{cb}^2 \, g_{bb}$, etc. The goal of our analysis
is to determine the allowed ranges of the six independent couplings.

\subsection{Additional observables}
\label{Addobs}

In addition to $2q2\ell$ operators, $VB$ exchange also produces
four-quark ($4q$) and four-lepton ($4\ell$) operators at tree
level. In the mass basis, the corresponding effective Lagrangian is
\bea
{\cal L}_\NP^{4q,4\ell} & = &
- \frac{g_{ij}^q g_{kl}^q}{2M^2_{V}} \(\oQ_{iL}\ga^\mu\si^I Q_{jL}\)
\(\oQ_{kL}\ga_\mu\si^I Q_{lL}\) \nn\\
&& \hskip2truecm
-~\frac{g_{ij}^\ell g_{kl}^\ell}{2M^2_{V}} \(\oL_{iL}\ga^\mu\si^I L_{jL}\)
\(\oL_{kL}\ga_\mu\si^I L_{lL}\) ~.
\label{4quark4lepton}
\eea
These contribute to five observables that yield important constraints
on the $VB$ model: $\bs$-$\bsbar$ mixing, neutrino trident production,
$\tau \to 3\mu$, $\tau \to \mu \nu {\bar\nu}$ and $D^0$-${\bar D}^0$
mixing. The first three have been discussed in detail in
Refs.~\cite{RKRDmodels,Alok:2017jgr}, the fourth in
Refs.~\cite{Isidori,Zurich}. The consideration of $D^0$-${\bar D}^0$
mixing is new. Below we summarize the constraints.

\subsubsection{$\bs$-$\bsbar$ mixing}

The SM contribution to $\bs$-$\bsbar$ mixing is generated via a box
diagram, and is given by
\beq
N C_{VLL}^{\rm SM} \, ({\bar s}_L \gamma^\mu b_L)\,({\bar s}_L \gamma_\mu b_L) ~.
\eeq
The operators of Eq.~(\ref{4quark4lepton}) include 
\beq
\frac{g_{sb}^2}{2 M^2_{V}} \, ({\bar s}_L \gamma^\mu b_L)\,({\bar s}_L \gamma_\mu b_L) ~,
\label{BsmixingVB}
\eeq
which generates a contribution to $\bs$-$\bsbar$ mixing.

Combining the SM and $VB$ contributions, we define
\beq
N C_{VLL} \equiv N C_{VLL}^{\rm SM} + \frac{g_{sb}^2}{2 M^2_{V}} ~,
\eeq
leading to
\begin{equation}
\Delta M_s =\frac{2}{3} m_{B_s} f_{B_s}^2 \hat B_{B_s}  \left | N C_{VLL} \right  | ~.
\end{equation}

Taking $f_{B_s}\sqrt{\hat B_{B_s}} = (266 \pm 18)$ MeV
\cite{Aoki:2016frl}, the SM prediction is 
\beq
\Delta M_s^{\rm SM} = (17.4 \pm 2.6)~{\rm ps}^{-1} ~.
\eeq
This is to be compared with the experimental measurement \cite{HFLAV}
\beq
\Delta M_s = (17.757 \pm 0.021)~{\rm ps}^{-1} ~.
\eeq
Treating the theoretical error as gaussian, this can be turned into a
constraint on $g_{sb}$ to be used in the fits:
\beq
\frac{g_{sb}}{M_V} = \pm (1.0^{+2.0}_{-3.9}) \times 10^{-3}~{\rm TeV}^{-1} ~.
\label{EQ:VB:Bsmixing}
\eeq

As was noted in Sec.~\ref{LQBsmix}, there are more recent calculations
of the hadronic parameters, and this is problematic for NP solutions
of the $\bsmumu$ anomalies, particularly the $Z'$
\cite{DiLuzio:2017fdq}.  However, these new values for the hadronic
parameters also cause problems for the SM itself, and so, as was done
in the case of the $U_1$ LQ model, we will await verification of this
new result before including it among the constraints.

\subsubsection{Neutrino trident production}

Neutrino trident production is the production of $\mu^+\mu^-$ pairs in
neutrino-nucleus scattering, $\nu_\mu N \to \nu_\mu N \mu^+
\mu^-$. The $Z'$ contributes to this process. Including both the SM
and NP contributions, the theoretical prediction for the cross section
is \cite{Alok:2017jgr}
\beq
\left. { \sigma_\text{SM+NP} \over \sigma_\text{SM} } \right|_{\nu N \to \nu N \mu^+ \mu^-}
=
\frac{1}{1+(1+4s_W^2)^2} \left [ \left ( 1 + \frac{v^2 g_{\mu\mu}^2}{M_V^2}  \right )^2
+ \left ( 1 +4 s_W^2 +  \frac{v^2 g_{\mu\mu}^2}{M_V^2}  \right )^2 \right ] ~.
\label{tridentHeavy}
\eeq
By comparing this with the experimental measurement \cite{CCFR}
\beq
 \left. { \sigma_\text{exp.} \over \sigma_\text{SM} } \right|_{\nu N \to \nu N \mu^+ \mu^-} = 0.82 \pm 0.28 ~,
\eeq
one obtains the following constraint on $g_{\mu\mu}$ to be used in the
fits:
\beq
\frac{g_{\mu\mu}}{M_V} = 0 \pm 1.13~{\rm TeV}^{-1} ~.
 \label{EQ:VB:trident}
\eeq

\subsubsection{$\tau \to 3\mu$}

The Lagrangian of Eq.~(\ref{4quark4lepton}) includes the operator
\beq
-\frac{g_{\mu\mu} \, g_{\mu\tau}}{2M^2_{V}} \, ({\bar \mu}_L \gamma^\mu \tau_L) \, ({\bar \mu}_L \gamma_\mu \mu_L) ~,
\label{tau3muamp}
\eeq
which generates the LFV decay $\tau\to 3 \mu$. Its decay rate is given
by
\beq
\cB(\tau^-\to\mu^-\mu^+\mu^-) = X \frac{(g_{\mu\mu} \, g_{\mu\tau})^2}{16 M^4_V}\frac{m^5_\tau\tau_\tau}{192\pi^3} ~,
\eeq
where $X \approx 0.94$ is a suppression factor due to the non-zero
muon mass \cite{RKRDmodels}.

At present, the experimental upper bound on the branching ratio for
this process is \cite{tau23muexp}:
\beq
\cB(\tau^-\to\mu^-\mu^+\mu^-) < 2.1 \times 10^{-8} ~ {\rm at~90\%~C.L.}~
\label{tau3muexp}
\eeq
This leads to
\beq
\frac{|g_{\mu\mu} \, g_{\mu\tau}|}{M^2_V} < 0.013~{\rm TeV}^{-2} ~.
 \label{EQ:VB:tau3mu}
\eeq
As we will see, when combined with the constraints from the $B$
anomalies and $\bs$-$\bsbar$ mixing, this puts an important bound on
$|g_{\mu\tau}/g_{\mu\mu}|$.

\subsubsection{$\tau \to \mu \nu {\bar\nu}$}

The Lagrangian of Eq.~(\ref{4quark4lepton}) also includes the operator
\beq
-\frac{1}{M^2_{V}} \, (- g_{\mu\tau} \, g_{ij} + 2 g_{\mu j} \, g_{i\tau})
\, ({\bar \mu}_L \gamma^\mu \tau_L) \, ({\bar \nu}_{iL} \gamma_\mu \nu_{jL}) ~,
\label{taumununubaramp}
\eeq
which generates the decay $\tau \to \mu \nu {\bar\nu}$.  The first
term in the coefficient is due to the tree-level exchange of a $Z'$,
while the second arises from $W'$ exchange. The SM also contributes to
this decay, but only for $i = \tau$ and $j = \mu$.  The decay rate is
then proportional to
\beq
\left\vert 1 + \frac{1}{2 \sqrt{2} G_F M_V^2} \, (- g_{\mu\tau}^2 + 2 g_{\mu\mu} \, g_{\tau\tau}) \right\vert^2
+ {\sum_{ij=\mu,\tau}}^{\hskip-1.6truemm \prime} \left\vert \frac{1}{2 \sqrt{2} G_F M_V^2} \, (- g_{\mu\tau} \, g_{ij} + 2 g_{\mu j} \, g_{i\tau}) \right\vert^2 ~,
\label{tmununubarrate}
\eeq
where the $\sum'_{ij=\mu,\tau}$ excludes $(i,j) = (\tau,\mu)$.

The most stringent constraint arises from an LFUV ratio of BRs.
However, a complication arises because there are two such ratios --
$\cB(\tau \to \mu \nu {\bar\nu}/\mu \to e \nu {\bar\nu})$ and
$\cB(\tau \to \mu \nu {\bar\nu}/\tau \to e \nu {\bar\nu})$ -- and
their measurements are not in complete agreement with one another
\cite{Pich:2013lsa}:
\bea
R_{\tau}^{\tau/e} &\equiv& \frac{\cB(\tau \to \mu \nu \bar \nu)/\cB(\tau \to \mu \nu \bar \nu)_{SM}}
{\cB(\mu \to e \nu \bar \nu)/\cB(\mu \to e \nu \bar \nu)_{SM}} = 1.0060 \pm 0.0030 ~, \\
R_{\tau}^{\mu/e} &\equiv& \frac{\cB(\tau \to \mu \nu \bar \nu)}{\xi_{ps} ~\cB(\tau \to e \nu \bar \nu)} 
= 1.0036 \pm 0.0028 ~,
 \label{EQ:VB:taumununubar}
\eea
where $\xi_{ps} = 0.9726$ is the phase-space factor. The first
measurement disagrees with the SM by $2\sigma$, while for the second
measurement, the disagreement is only at the level of $1.3\sigma$.
Both of these apply to the quantity in Eq.~(\ref{tmununubarrate}), and
we include both constraints in the fits.

As we will see, $g_{\mu\tau}$ is quite small in the $VB$ model. If it
is neglected in Eq.~(\ref{tmununubarrate}), one obtains the constraint
\beq
\frac{|g_{\mu\mu} \, g_{\tau\tau}|}{M^2_V} =
\begin{cases}
      0.049 \pm 0.025 ~{\rm TeV}^{-2} ~, & R_{\tau}^{\tau/e} \\
      0.030 \pm 0.023 ~{\rm TeV}^{-2} ~, & R_{\tau}^{\mu/e}
    \end{cases}
 \label{EQ:VB:gmumugtautau}
\eeq
Conservatively, this gives $|g_{\mu\mu} \, g_{\tau\tau}|/M^2_V < 0.1 ~{\rm TeV}^{-2}$.

\subsubsection{$D^0$-${\bar D}^0$ mixing}

$D^0$-${\bar D}^0$ mixing has been measured experimentally. It is
found that \cite{pdg}
\beq
\Delta M_D = (0.95^{+0.41}_{-0.44}) \times 10^{-2} ~{\rm ps}^{-1} ~.
\label{EQ:data}
\eeq
Within the SM there are two types of contributions to $D^0$-${\bar
  D}^0$ mixing. The short-distance contributions, governed by the
quark-level box diagrams, yield $\Delta M_D = O(10^{-4})~{\rm
  ps}^{-1}$ \cite{Barger:1989fj}, too small to explain the data. The
long-distance contribution, from hadron exchange, is estimated to be
$\Delta M_D = (1$-$46) \times 10^{-3}~{\rm ps}^{-1}$
\cite{Barger:1989fj}. Thus, it can account for the measured value of
$\Delta M_D$, though the estimate is very uncertain.

We therefore see that $\Delta M_D$ receives both short- and
long-distance contributions, but the latter are difficult to compute
with any precision. Thus, constraints on any NP contributions should
really focus on the short-distance effects. In
Ref.~\cite{Bevan:2012waa}, all available data have been combined to
extract the fundamental quantities $|M_{12}|$ and $|\Gamma_{12}|$.
Their fit yields
\beq
 |M_{12}|^{\rm data} = (6.9 \pm 2.4) \times 10^{-3}\, {\rm ps}^{-1} ~~,~~~~
 |\Gamma_{12}|^{\rm data} = (17.2 \pm 2.5) \times 10^{-3}\, {\rm ps}^{-1} \,. 
 \label{EQ:data}
\eeq
$|M_{12}|^{\rm data}$ will be used to constrain the NP.

In the $VB$ model, there is a contribution to $D^0$-${\bar D}^0$
mixing from the tree-level exchange of the $Z'$.  We have
\beq
 H_{\rm eff}^{Z'} = \frac{(g_{uc})^2}{2M_V^2}  ({\bar c}_L \gamma_\mu u_L) \, ({\bar c}_L \gamma^\mu u_L) ~, 
\eeq
where $g_{uc}$ is the ${\bar c}_L u_L Z'$ coupling. This leads
to
\beq
 |M_{12}|^{Z'} = \frac13 m_D f_D^2 \hat B_D \left| \frac{(g_{uc})^2}{2M_V^2} \right|^2 ~.
\eeq
To be conservative, we require only that $|M_{12}|^{Z'}$ be less than
the experimental measurement of Eq.~(\ref{EQ:data}). Taking $f_D =
(212.15 \pm 1.45)$ MeV and $\hat B_D = 0.75 \pm 0.03$
\cite{Aoki:2016frl}, this leads to
\beq
|g_{uc}| \le 6.6 \times 10^{-4} ~ (M_V / 1~{\rm TeV}) ~.
\label{guclimit}
\eeq

Now,
\bea
g_{uc} & = & V_{cb} V_{ub}^* g_{bb} + (V_{cs}V_{ub}^*+V_{cb}V_{us}^*) g_{sb} + V_{cs}V_{us}^* g_{ss} \nn\\
 & \simeq & (0.5 + 1.3 i) \times 10^{-4} g_{bb} + (1.0 + 0.3 i) \times 10^{-2} g_{sb} + 0.22\, g_{ss} ~.
\eea
(Note that, although $g_{bb}$, $g_{sb}$ and $g_{ss}$ are real,
$g_{uc}$ is complex due to the CKM matrix elements.) Applying the
constraint of Eq.~(\ref{guclimit}) to each of the terms individually,
we find
\beq
|g_{bb}| \le 4.7 ~ (M_V / 1~{\rm TeV}) ~~,~~~~ 
|g_{sb}| \le 6.3 \times 10^{-2} ~ (M_V / 1~{\rm TeV}) ~~,~~~~ 
|g_{ss}| \le 3 \times 10^{-3} ~ (M_V / 1~{\rm TeV}) ~.
\label{gssconstraint}
\eeq
Now, for $M_V = 1$ TeV, $|g_{bb}| \le 1$ has been imposed for
perturbativity, and $|g_{sb}| \le O(10^{-3})$
[Eq.~(\ref{EQ:VB:Bsmixing})], so the above bounds are irrelevant for
these couplings. However, the bound on $|g_{ss}|$ is important since
it is the only constraint on this coupling.

\subsection{Fits}
\label{Sec:VBFits}

The $VB$ model contributes at tree level to a large number of
observables, resulting in 15 constraints that must be included in the
fit (we do not consider the RGE constraints). They are found in
Table~\ref{tab:obs_meas} ($2q2\ell$ observables, 11 constraints),
Eq.~\eqref{EQ:VB:Bsmixing} ($4q$, 1) and Eqs.~\eqref{EQ:VB:trident}
and \eqref{EQ:VB:gmumugtautau} ($4\ell$, 3). In addition, the
condition of Eq.~\eqref{EQ:VB:tau3mu} must be taken into account. We
now perform a fit in which the 6 couplings are the unknown parameters
to be determined.

Before presenting the results of the fit, it is a very useful exercise
to deduce the general pattern of the values of the couplings
(throughout, $M_V = 1$ TeV is assumed):
\begin{enumerate}

\item The constraint from $\bs$-$\bsbar$ mixing requires $|g_{sb}|
  \lsim O(10^{-3})$ [Eq.~(\ref{EQ:VB:Bsmixing})].

\item $C_9 = -C_{10}$ is proportional to $g_{sb} g_{\mu\mu}$. The
  constraint from the $b \to s \mu\mu$ data leads to $g_{sb}g_{\mu\mu}
  = -0.0011 \pm 0.0002$. Since $|g_{sb}| \lsim
  O(10^{-3})$, this then imples that $g_{\mu\mu} \lsim O(1)$.

\item The constraint from $\tau \to 3\mu$ [Eq.~(\ref{EQ:VB:tau3mu})]
  requires $|g_{\mu\mu} \, g_{\mu\tau}| < 0.013$. Given that
  $g_{\mu\mu} \lsim O(1)$, this implies that $|g_{\mu\tau}/g_{\mu\mu}|
  \ll 1$.

\item Since $g_{\mu\tau}$ is very small, the constraint of
  Eq.~(\ref{EQ:VB:gmumugtautau}) applies. And since $g_{\mu\mu} \lsim
  O(1)$, this leads to $|g_{\tau\tau}|$ in the range 0.01-0.1.

\item $C_V$ is proportional to $(V_{cs} \, g_{sb} + V_{cb} \, g_{bb})
  g_{\tau\tau}$. The constraint from the $R_{D^{(*)}}$ anomaly implies
  that $(V_{cs} \, g_{sb} + V_{cb} \, g_{bb}) g_{\tau\tau} = 0.07 \pm
  0.02$. Since $|g_{sb}| \lsim O(10^{-3})$, the first term is
  negligible, so that the $V_{cb} \, g_{bb} g_{\tau\tau}$ term
  dominates. (This is opposite to the $U_1$ LQ, where the first term
  dominated.)

\item $C_L$ is proportional to $g_{sb}(g_{\mu\mu} + g_{\tau\tau})$. In
  order to evade the constraint from $B \to K^{(*)} \nu {\bar\nu}$, we
  require $-0.014 \le g_{sb}(g_{\mu\mu} + g_{\tau\tau}) \le 0.034$.
  However, because $|g_{sb}| \lsim O(10^{-3})$, this is always
  satisfied, so there are no additional constraints on the couplings
  from this process.

\item Above we found $|g_{\mu\tau}/g_{\mu\mu}| \ll 1$. For such small
  values of $g_{\mu\tau}$, there are no constraints from the
  semileptonic LFV decays. 

\item The only constraint on $g_{ss}$ is in Eq.~(\ref{gssconstraint}):
  $|g_{ss}| \le 3 \times 10^{-3}$.

\end{enumerate}
The key point is \#5 above. Recall that $R_{D^{(*)}}^{\tau/\ell} =
{\cal B}(B^- \to D^{(*)} \tau^- {\bar\nu}_\tau)/{\cal B}(B^- \to
D^{(*)} \ell^- {\bar\nu}_\tau)$ $\ell = e, \mu$. Assuming that the NP
affects mainly $B^- \to D^{(*)} \tau^- {\bar\nu}_\tau$, in order to
reproduce the measured values of $R_{D^{(*)}}$, we require both
$g_{bb}$ and $g_{\tau\tau}$ to be large, $O(1)$. However, from \#4, we
see that $g_{\tau\tau}$ is constrained to be quite a bit smaller. In
light of this, the NP contribution to the $\bctaunu$ amplitude is also
small. The only way to generate an enhancement of $R_{D^{(*)}}$ is if
the amplitudes in the denominator are suppressed. Now, the NP can
affect only $B^- \to D^{(*)} \mu^- {\bar\nu}_\mu$, with a contribution
proportional to $V_{cb} \, g_{bb} \, g_{\mu\mu}$. Since both $g_{bb}$
and $g_{\mu\mu}$ are $O(1)$, this contribution can be important,
leading to a suppression only if $g_{bb} \, g_{\mu\mu} < 0$.  On the
other hand, if such an effect were present, it would lead to a large
value of $R_{D^*}^{e/\mu}/(R_{D^*}^{e/\mu})_\SM$, and this is not
observed (see Table \ref{tab:obs_meas}). This constraint limits the
size of the NP contribution to $B^- \to D^{(*)} \mu^- {\bar\nu}_\mu$.
The bottom line is that, while this general $VB$ model can lead to an
enhancement of $R_{D^{(*)}}$ over its SM values, it cannot reproduce
the measured central values of $R_{D^{(*)}}$. This will necessarily
increase the $\chi^2$ of the fit.

After performing the fit, we find $\chi_{min,\SM+VB}^2 = 15$. Since
the d.o.f.\ is 10 (15 constraints, 5 independent couplings, since we
have only a single constraint on $g_{ss}$), this gives
$\chi^2_{min}/{\rm d.o.f.} = 1.5$, which is a marginal fit. But we
understand where the problem lies: the $VB$ model cannot explain the
measured central values of $R_{D^{(*)}}$. In fact, the typical value
of $R_{D^{(*)}}$ that is generated in this model is roughly $2\sigma$
below the measured values. As such, the observables
$R_{D^*}^{\tau/\ell}/(R_{D^*}^{\tau/\ell})_\SM$ and
$R_{D}^{\tau/\ell}/(R_{D}^{\tau/\ell})_\SM$ contribute $\chi^2 \sim 8$
by themselves to $\chi_{min,\SM+VB}^2$.

Even so, we do not feel that this $VB$ model should be discarded.
After all, it {\it can} simultaneously explain anomalies in $\bsmumu$
and $\bctaunu$ transitions. Obviously, if future measurements of
$R_{D^{(*)}}$ confirm the present size of the discrepancy with the SM,
the $VB$ model will be excluded. However, if it turns out that the
central values of $R_{D^{(*)}}$ are reduced, the $VB$ model will be as
viable an explanation as the $U_1$ LQ model. For this reason, in what
follows we refer to this $VB$ model as semi-viable.

The best-fit values of the couplings are 
\bea
& g_{\mu\mu} = -0.95^{+0.42}_{-0.72} ~~,~~~~ g_{\mu\tau} = 0.0 \pm 0.018 ~~,~~~~ g_{\tau\tau} = -0.039^{+0.019}_{-0.037} ~, & \nn\\
& g_{bb} = 0.85^{+0.96}_{-0.41} ~~,~~~~ g_{sb} = (1.1^{+0.9}_{-0.2}) \times 10^{-3} ~~,~~~~ |g_{ss}| \le 3 \times 10^{-3} ~, &
\label{VBfitresults}
\eea
for $M_V = 1$ TeV.  We have several observations. First, as was the
case with the $U_1$ LQ model (Sec.~\ref{U1LQfit}), the couplings are
very poorly determined in the fit. This is again because the
observables depend almost exclusively on products of the couplings,
and so yield only imprecise information about the individual
couplings. Even so, these values and errors indicate the size of the
couplings, and these agree with our rough estimates above.

Second, we note that $g_{\mu\tau}$ is quite small. Indeed, after
performing a scan over the parameter space, we find that
$|g_{\mu\tau}/g_{\mu\mu}| \le 0.1$ (95\% C.L.). Now, the two previous
analyses \cite{RKRDmodels,Zurich} made the assumption that the NP
couples predominantly to the third generation. The couplings involving
the second generation obey a hierarchy $|c_{22}| < |c_{23}|, |c_{32}|
< |c_{33}|$, where the indices indicate the generations. To be
specific, these analyses have $|g_{\mu\tau}/g_{\mu\mu}| > 1$. But this
is in clear disagreement with the data, so that the $VB$ model is
excluded as an explanation of the $\bsmumu$ and $\bctaunu$ anomalies.
On the other hand, as we have seen above, the general $VB$ model {\it
  is} semi-viable. We therefore conclude that its exclusion by the
previous analyses is directly due to their theoretical assumption
about the NP couplings.

In the interest of accuracy, it must be said that this was not the
argument used by previous analyses to exclude the $VB$ model. For
example, in Ref.~\cite{Zurich}, the breaking of the $U(2)_q \times
U(2)_\ell$ flavour symmetry led naturally to values of $O(0.1)$ for
$g_{sb}$. (This in turn implies a small value for $g_{\mu\mu}$. With
$g_{\mu\tau} \simeq 0.1$ and $g_{\mu\mu} \simeq 0.01$, the authors
found that the constraint from $\tau\to 3\mu$ was satisfied [we agree,
  see Eq.~(\ref{EQ:VB:tau3mu})].) Of course, such large values of
$g_{sb}$ are in conflict with the constraints from $\bs$-$\bsbar$
mixing [Eq.~(\ref{EQ:VB:Bsmixing})]. However, Ref.~\cite{Zurich}
focused on the $2q2\ell$ observables, and found that the $B \to
K^{(*)} \nu {\bar\nu}$ and RGE constraints ruled out the $VB$ model.

Above, we found values for the couplings of the general $VB$ model
that render it semi-viable. We would like to understand the origin
of this pattern of couplings. As we have seen, $g_{\mu\tau}$ is
predicted to be very small. Ideally, we would like a small value of
$g_{\mu\tau}$ to be the result of a symmetry. Now, it is often
asserted that, if a model violates lepton flavour universality, it
will also lead to lepton flavour violation\footnote{This was the main
  point of Ref.~\cite{GGL}. To illustrate this, the scenario of NP
  that couples only to the third generation in the weak basis was
  used.}. However, this is not necessarily true. In Ref.~\cite{AGC},
it is pointed out that it is possible to construct models that violate
LFU, but do not lead to sizeable LFV. This occurs when Minimal Flavour
Violation \cite{Chivukula:1987py, DAmbrosio:2002vsn,
  Cirigliano:2005ck, Davidson:2006bd, Alonso:2011jd} is used to
construct the model. Perhaps this $VB$ model is of this type.

\subsection{LHC Constraints}
\label{Sec:LHCConstraints}

ATLAS and CMS have examined $pp$ collisions at $\sqrt{s}$ = 13 TeV and
searched for high-mass resonances decaying into lepton pairs
\cite{Aaboud:2016cth,CMS:2016abv}. In the $VB$ model, it is a
reasonable approximation to consider only the $Z'$ couplings to
$b{\bar b}$ and $\mu^+\mu^-$ [see Eq.~(\ref{VBfitresults})]. In this
case, the relevant process is $b{\bar b} \to Z' \to \mu^+\mu^-$. Using
this process, and assuming $M_{Z'} = 1$ TeV, the non-observation of
resonances at the LHC puts the following constraint on the couplings:
\beq
\frac{1.1~g_{bb}^2 g_{\mu\mu}^2}{6.0~g_{bb}^2 + 2~g_{\mu\mu}^2} \le 3.1 \times 10^{-3} ~~~(95\%~{\rm C.L.})
\label{LHClimit}
\eeq

\begin{figure}[h]
\begin{center}
\includegraphics[width=0.4\textwidth]{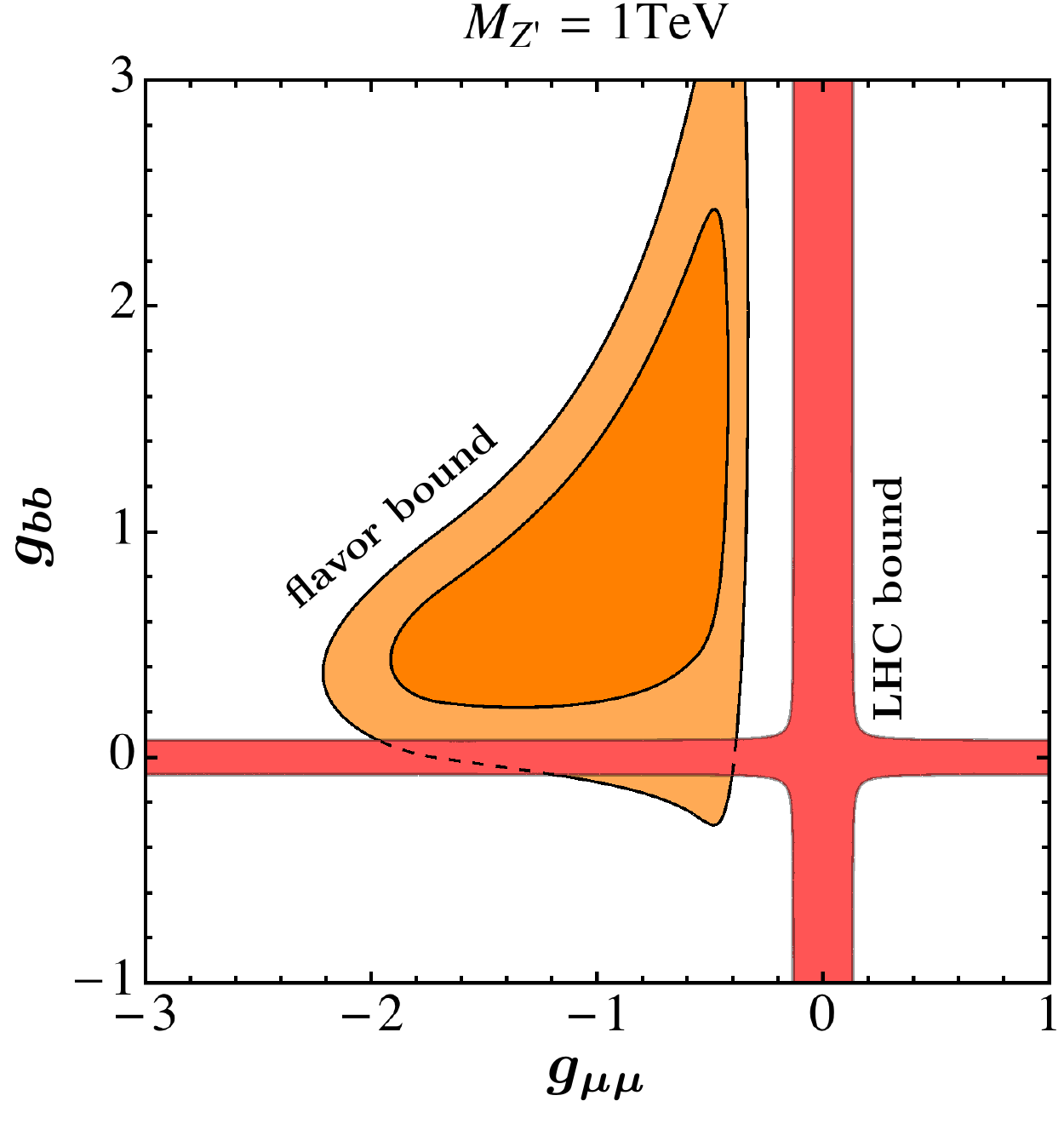}
\caption{Allowed regions in $(g_{bb},g_{\mu\mu})$ space from flavour
  and LHC constraints, assuming $M_{Z'} = 1$ TeV. The $1\sigma$ and
  $2\sigma$ flavour bounds are shown respectively in the dark and
  light orange regions. The 95\% C.L.\ LHC bound is shown in the red
  region.}
\label{LHCdimuon}
\end{center}
\end{figure}

In Fig.~\ref{LHCdimuon} we show the allowed regions in
$(g_{bb},g_{\mu\mu})$ space from the flavour and LHC constraints. At
$1\sigma$ in the flavour constraints, the regions do not overlap.
However, the $2\sigma$ region does overlap, suggesting that the VB
model might be viable. In order to quantify this, we include the LHC
result in the fit by converting the 95\% C.L. upper limit (UL) of
Eq.~(\ref{LHClimit}) to a bound of $0 \pm {\rm UL}/2$:
\beq
\frac{1.1~g_{bb}^2 g_{\mu\mu}^2}{6.0~g_{bb}^2 + 2~g_{\mu\mu}^2} = 0.0 \pm 1.55 \times 10^{-3} ~.
\label{LHClimit}
\eeq
Now we find $\chi_{min,\SM+VB}^2 = 19.3$.  The d.o.f.\ is 11 (16
constraints, 5 independent couplings), so that $\chi^2_{min}/{\rm
  d.o.f.} = 1.8$, which is a poor fit. 

We are forced to conclude that, in the end, the $VB$ model is excluded
as a possible combined explanation of the $\bsmumu$ and $\bctaunu$
anomalies. We stress that this conclusion is independent of any
assumption about the NP couplings. It is found simply by taking into
account all the flavour constraints and the bound from the LHC dimuon
search.

There is one possible loophole. If the $Z'$ has additional, invisible
decays, perhaps to dark matter \cite{Cline:2017lvv}, the LHC
constraints can be evaded. In this case, the $VB$ model would still be
permitted.

\section{Conclusions}
\label{Sec:Conclusions}

At the present time, there are a number of measurements that are in
disagreement with the predictions of the SM. The observables all
involve the quark-level transitions $\bsmumu$ or $\bctaunu$. It was
shown that, theoretically, both anomalies could be explained within
the same new-physics model, and four possibilities were identified.
There are three leptoquark models -- $S_3$, $U_3$, $U_1$ -- and the
$VB$ model, containing SM-like $W'$ and $Z'$ vector bosons. These four
NP models were examined in recent analyses, under the theoretical
assumption that the NP couples predominantly to the third generation,
with the couplings involving the second generation subdominant. It was
found that, when constraints from other processes are taken into
account, the $S_3$, $U_3$ and $VB$ models cannot explain the $B$
anomalies, but $U_1$ is viable. However, this raises the question: to
what extent do these conclusions depend on the theoretical assumption
regarding the NP couplings? In this paper, we reanalyze the models,
but without any assumption about their couplings.

In LQ models, there are new tree-level contributions to semileptonic
processes involving two quarks and two leptons.  Now, several of the
$B$ anomalies violate lepton flavour universality, suggesting that any
NP explanations may also lead to lepton-flavour-violating effects. And
indeed, there are several $2q2\ell$ LFV processes: $B \to K^{(*)}
\mu^\pm \tau^\mp$, $\tau \to \mu \phi$, $\Upsilon \to \mu\tau$ and
$J/\psi \to \mu \tau$. However, these were not fully taken into
account in previous analyses. We find that constraints from these
processes are extremely important.

For the LQ models, we show that, even if the LFV constraints are not
applied, $S_3$ and $U_3$ cannot explain the $B$-decay anomalies. On
the other hand, $U_1$ is a viable model. The problem is that, while
products of the LQ couplings are found to lie in certain ranges of
values, there is very little information about the individual
couplings themselves. This is greatly improved when the LFV
constraints are added. We find that the region of allowed couplings is
greatly reduced, and is similar (though somewhat larger) to that found
when the NP couples predominantly to the third generation.  That is,
the experimental data suggest a pattern of LQ couplings similar to
that of the theoretical assumption.

The LFV constraints have an additional effect. The scale of NP is well
above the weak scale. When the NP is integrated out, this produces
${\cal L}_\NP$, which contains effective four-fermion operators. It is
assumed that these are dominated by the $2q2\ell$ operators that
contribute to $\bsmumu$ or $\bctaunu$. When the full Lagrangian,
${\cal L}_\SM + {\cal L}_\NP$, is evolved to low energies using the
renormalization group equations, this produces new operators and
corrections to SM operators. It has been argued that all of these
effects lead to additional, important constraints on the NP, and
reduce the region of allowed couplings. In this paper, we point out
that these constraints are not rigorous. In real models, ${\cal
  L}_\NP$ may contain additional operators, both dominant and
subdominant, that can change the conclusions of the RGE analysis.  But
even if one accepts the RGE constraints, we show that, if one requires
$|h_{ij}| \le 1$ (so that the couplings remain perturbative), the LFV
constraints, which were ignored in the RGE discussion, lead to a
much larger reduction of the allowed region of NP couplings. That is,
the RGE constraints are unimportant.

The $U_1$ LQ model is therefore a viable candidate for simultaneously
explaining the $\bsmumu$ or $\bctaunu$ anomalies. If correct,
observable effects in other processes are predicted. In particular,
the violation of lepton flavour universality in $B \to \pi \ell
{\bar\nu}_\ell$ or $B^- \to \ell {\bar\nu}_\ell$ decays may be
enhanced over the SM by as much as 40\%.  $\cB(B \to K^{(*)} \nu
{\bar\nu})$ may be increased by 70\% over the SM. And $\cB(B \to
K^{(*)} \tau^+ \tau^-)$ and ${\cal B}(\bs \to \tau^+ \tau^-)$ may be
enhanced by as much as three orders of magnitude. Most importantly,
these predictions are correlated with one another, and with the value
of $R_{D^*}^{\tau/\ell}/(R_{D^*}^{\tau/\ell})_\SM$. This is a good
test of the $U_1$ model.

For the $VB$ model, the conclusions are quite different than for the
LQ models. First, there are also tree-level contributions to
four-quark and four-lepton observables, and these lead to important
additional constraints on the couplings (values are given assuming
$M_V = 1$ TeV). In particular, the constraint from $\bs$-$\bsbar$
mixing implies that $|g_{sb}| \lsim O(10^{-3})$.  In turn, in order to
explain the $\bsmumu$ anomaly, $g_{\mu\mu} \lsim O(1)$ is
required. Finally, in order to evade the constraint from $\tau\to
3\mu$, $g_{\mu\tau}$ must be sufficiently small. We find that, when
all constraints are applied to the $VB$ model,
$|g_{\mu\tau}/g_{\mu\mu}| < 0.1$ is required. If the NP couples
predominantly to the third generation, it is found that the $Z'$
couplings involving the second-generation leptons obey $|g_{\mu\tau}|
> |g_{\mu\mu}|$.  This clearly rules out the $VB$ model with the above
theoretical assumption about its couplings. (Previous analyses also
ruled out the $VB$ model, but for other reasons.)

Another process to which $VB$ contributes at tree level is $\tau \to
\mu \nu {\bar\nu}$, and the constraints are very stringent. Given that
$g_{\mu\mu} \lsim O(1)$, they lead to a value for $|g_{\tau\tau}|$ in
the range 0.01-0.1. The NP contribution to $B^- \to D^{(*)} \tau^-
{\bar\nu}_\tau$ is proportional to $g_{bb} \, g_{\tau\tau}$.  Even
though $g_{bb} = O(1)$, such a small value of $|g_{\tau\tau}|$ leads
to a small NP effect, and makes it impossible to reproduce the
measured central values of $R_{D^{(*)}}$. There is an enhancement of
$R_{D^{(*)}}$ (due to a suppression of $B^- \to D^{(*)} \mu^-
{\bar\nu}_\mu$), but it is smaller than what is observed.  Thus, the
$VB$ model would be viable only if future measurements find that the
central values of $R_{D^{(*)}}$ are reduced.

Now, the process $b{\bar b} \to Z' \to \mu^+\mu^-$ leads to the
production of high-mass resonant dimuon pairs in $pp$ collisions at
$\sqrt{s}$ = 13 TeV.  Unfortunately, since both $g_{bb}$ and
$g_{\mu\mu}$ are $\lsim O(1)$, this leads to a production rate larger
than the limits placed by ATLAS and CMS. The only way to evade this is
if the $Z'$ has additional, invisible decays. If this does not occur,
the upshot is that, in the end, the $VB$ model is excluded as a
possible combined explanation of the $B$ anomalies.  However, this
conclusion is not the result of any assumption about the NP
couplings. Rather, it is found simply by taking into account all the
flavour constraints and the bound from the LHC dimuon search.

\bigskip
\noindent
{\bf Acknowledgments}: JK would like to thank David Straub and Michael
Paraskevas for very useful discussions.  This work was financially
supported by NSERC of Canada (DL, RW).

\end{document}